\documentclass[fleqn,usenatbib]{mnras}
\usepackage{newtxtext,newtxmath}
\usepackage{comment}
\usepackage[T1]{fontenc}
\DeclareRobustCommand{\VAN}[3]{#2}
\let\VANthebibliography\thebibliography
\def\thebibliography{\DeclareRobustCommand{\VAN}[3]{##3}\VANthebibliography}
\usepackage{graphicx}	
\usepackage{amsmath}	
\usepackage{bigints}
\usepackage{threeparttable}
\usepackage{scalerel,tikz}
\usepackage{arydshln}
\usepackage{booktabs}
\usetikzlibrary{svg.path}
\definecolor{orcidlogocol}{HTML}{A6CE39}
\tikzset{orcidlogo/.pic={
 \fill[orcidlogocol] svg{M256,128c0,70.7-57.3,128-128,128C57.3,256,0,198.7,0,128C0,57.3,57.3,0,128,0C198.7,0,256,57.3,256,128z};
 \fill[white] svg{M86.3,186.2H70.9V79.1h15.4v48.4V186.2z}
 svg{M108.9,79.1h41.6c39.6,0,57,28.3,57,53.6c0,27.5-21.5,53.6-56.8,53.6h-41.8V79.1z M124.3,172.4h24.5c34.9,0,42.9-26.5,42.9-39.7c0-21.5-13.7-39.7-43.7-39.7h-23.7V172.4z}
 svg{M88.7,56.8c0,5.5-4.5,10.1-10.1,10.1c-5.6,0-10.1-4.6-10.1-10.1c0-5.6,4.5-10.1,10.1-10.1C84.2,46.7,88.7,51.3,88.7,56.8z};
}}
\newcommand\orcidicon[1]{\href{https://orcid.org/#1}{\mbox{\scalerel*{
\begin{tikzpicture}[yscale=-1,transform shape]
\pic{orcidlogo};
\end{tikzpicture}
}{|}}}}
\newcommand{\tikzxmark}{%
\tikz[scale=0.23] {
    \draw[line width=0.7,line cap=round] (0,0) to [bend left=6] (1,1);
    \draw[line width=0.7,line cap=round] (0.2,0.95) to [bend right=3] (0.8,0.05);
}}

\newcommand{\rits}[1]{{\color{black}#1}} 
\newcommand{\maxg}[1]{{\color{black}#1}} 

\title[Cloud crushing with driven turbulence]{Woven by the Whirls: The growth and entrainment of cold clouds in turbulent hot winds}
\author[Ghosh et al.]{
Ritali Ghosh $^{\orcidicon{0000-0001-8643-7104}}$,$^{1,2}$\thanks{E-mail: rgosh@mpa-garching.mpg.de}
Max Gronke $^{\orcidicon{0000-0003-2491-060X}}$,$^{3,1}$ 
Prateek Sharma $^{\orcidicon{0000-0003-2635-4643}2}$ and
Alankar Dutta $^{\orcidicon{0000-0002-9287-4033}1}$
\\
$^{1}$ Max Planck Institute for Astrophysics, Garching 85748, Germany\\
$^{2}$ Joint Astronomy Programme, Department of Physics, Indian Institute of Science, Bangalore 560012, India\\
$^{3}$ Astronomisches Rechen-Institut, Zentrum für Astronomie, Universität Heidelberg, Mönchhofstraße 12-14, 69120 Heidelberg, Germany
}

\date{Accepted XXX. Received YYY; in original form ZZZ}

\pubyear{2025}

\begin{document}
\label{firstpage}
\pagerange{\pageref{firstpage}--\pageref{lastpage}}
\maketitle

\begin{abstract}
Galactic and intergalactic flows often exhibit relative motion between the cold dense gas and the hot diffuse medium.
Such multiphase flows -- involving gas at different temperatures, densities, and ionization states -- for instance, galactic winds, are frequently turbulent. 
However, idealized simulations typically model the winds and driven turbulence separately, despite their intertwined roles in galaxy evolution. To address this, we investigate the survival of a dense cloud in a hot wind subject to continuous external turbulent forcing. We perform 3D hydrodynamic simulations across a range of turbulent Mach numbers in the hot phase $\mathcal{M}_{\rm turb}=v_{\rm turb}/c_{\rm s, wind}$ from 0.1 to 0.7 ($c_{\rm s, wind}$ and $v_{\rm turb}$ being the sound speed and the turbulent velocity in the hot phase, respectively).
We find that in spite of the additional subsonic turbulence, cold clouds can survive if the cooling time of the mixed gas \rits{$t_{\rm cool, mix}$} is shorter than a modified destruction time \rits{$\tilde{t}_{\rm cc}$}, i.e., 
$t_{\rm cool,mix}/\tilde{t}_{\rm cc}<1$ where \rits{$\tilde{t}_{\rm cc}=t_{\rm cc}/(1+\left(\mathcal{M}_{\rm turb}/\left(f_{\rm mix}\mathcal{M}_{\rm wind}\right)\right)^2)^{1/2}$, where $f_{\rm mix}\sim0.6$ is a fudge factor.}
Moreover, in the `survival regime', turbulence can enhance the growth of cold clouds by up to an order of magnitude because of more efficient stretching and an associated increase in the surface area.
This increase in mass transfer between the phases leads to significantly faster entrainment of cold material in turbulent winds. 
In contrast to the narrow filamentary tails formed in laminar winds, turbulence stretches the cold gas orthogonally, dispersing it over a larger area and changing absorption line signatures.
\end{abstract}

\begin{keywords}
galaxies: evolution -- galaxies: haloes -- hydrodynamics -- galaxies: kinematics and dynamics
\end{keywords}

\section{Introduction}\label{sec:intro}

Multiphase flows \rits{-- characterized by a wide} 
range of densities, temperatures, and ionization states -- are routinely observed in diverse astrophysical systems across different scales. 
\rits{Notable examples include} the interstellar medium (ISM; \citealt{BegelmanFabian1990MNRAS.244P..26B, Heitsch2008, Audit2010}), the circumgalactic medium (CGM; \citealt{Tumlinson/annurev-astro-091916-055240, Heckman_2017}), the intracluster medium (ICM;  \citealt{Bregman_2006, Olivares2019}), and even the solar corona (\citealt{Antolin_2020, Hillier2023MNRAS}). 
\rits{Within galactic environments, such flows manifest in various forms, including} galactic outflows driven by stellar feedback (\citealt{Veilleux2005, Bordoloi2014ApJ, Rubin2014ApJ, McQuinn2019ApJ}), the motion of HI-rich high-velocity clouds (HVCs) towards the Galactic center (\citealt{Muller1963, Putman2003, Fox_2006, Richter2017A&A}), the spectacular tails of jellyfish galaxies moving through the ICM (\citealt{Sun_2010, Fumagalli2014, Poggianti_2019}), the Magellanic Stream in our Milky Way (\citealt{Putman2003, Fox2014, Krishnarao2022Nature}), the H$\alpha$ filaments in cool-core clusters (\citealt{Heckman1989,McDonald2012ApJ, Lakhchaura2018}), and the extended HI emission and absorption along radio jets (\citealt{Morganti2010}).
\rits{At the core of these diverse flows lies the same physical process: the interaction of cold ($\lesssim10^4$ K) and dense gas with a hot ($\gtrsim 10^6$ K) and diffuse medium.}
Understanding the evolution of cold, dense clouds embedded in hot, diffuse wind is therefore of great relevance in the formation and evolution of galaxies. 

\rits{The interaction between cold and hot gas phases is a complex multi-physics problem involving}
\rits{shear at the interface} between hot 
and cold gas,  
fast radiative cooling of warm ($\sim 10^5$ K) \rits{phase formed by mixing, 
as well as magnetic fields, thermal conduction and turbulence within boundary layers.} 
Several recent works have focused on various combinations of these fundamental processes and obtained valuable insights into the physics of such multiphase flows. Three idealized simulation setups have been extensively explored: the radiative cloud-crushing problem \citep{Armillotta2017,GronkeOh2018MNRAS.480L.111G, Gronke2020, Sparre2020MNRAS, Kanjilal2021, Gronnow2021, Girichidis2021MNRAS,Abruzzo2022ApJ}, 
the radiative mixing layer \citep{Kwak2010, Ji2019, Fielding2020ApJ, TanOhGronke2021, Prateek2025arXiv}, 
and the multiphase turbulent box (thermal instability with cold gas condensing out of a turbulent hot phase in global thermal equilibrium: \citealp{Mohapatra2019, Grete_2020,Mohapatra2022MNRAS}; cold gas subject to turbulent forcing but with net radiative cooling: \citealp{Gronke2022, Das2024MNRAS}). All of these have been studied with and without magnetic fields and (anisotropic-) thermal conduction, which do not seem to qualitatively alter the outcome of these simulations. 
\rits{
However, the cold and hot gas interaction is inherently turbulent due to the large Reynolds numbers typical in astrophysical systems. The stochastic and spatially inhomogeneous nature of mass and energy injection from processes such as star formation and AGN activity, as well as the motion of galactic substructures drives turbulence in the circumgalactic and intracluster media. 
}
To 
\rits{explore the effect of this ubiquitous turbulence on the interaction between cold clouds and hot wind}, in this work, we introduce externally driven turbulence 
in the idealized problem of radiative cloud-crushing. 

The standard \textit{cloud-crushing} simulations \rits{are designed to study} the evolution of a cold and dense blob, initially at rest, subjected to a hot laminar wind \rits{
-- an idealization of cold clumps uplifted from the ISM and exposed to a hot galactic outflow.} 
\rits{Several} studies suggest that, in the presence of radiative cooling, clouds larger than a threshold size can grow by continuous cooling of the mixed gas with a short cooling time (\citealt{GronkeOh2018MNRAS.480L.111G, Armillotta2017, Kanjilal2021, Abruzzo2022ApJ}). The overall result is that the mass from the hot wind condenses on to the growing cold cometary tail. These works highlight the importance of the timescale of radiative cooling of the mixed gas - produced by shear between the hot and cold phases - in determining the survival and growth of cold gas content in outflows. A shorter cooling time for mixed gas $t_{\rm cool, mix}$ compared to the hydrodynamic destruction time $t_{\rm cc}$ leads to a continuous increase in the mass of cold gas. 
\rits{Since $t_{\rm cc}$ scales with cloud size, this condition translates into a threshold size:}
clouds \rits{larger than a critical size} 
can survive. 

The \textit{radiative mixing layer} simulations (\citealt{Esquivel2006ApJ,Kwak2010, 
Ji2019, Fielding2020ApJ, TanOhGronke2021, Prateek2025arXiv}) complement the above picture by resolving the interface between the hot and cold phases in a shear flow setup. In these simulations, hot gas flowing into the radiative mixing layer (where mixed gas at intermediate temperatures cools rapidly) brings in mass, enthalpy (most of which is radiative away), and momentum to the cold phase. 
This funneling of the hot gas into the radiative mixing layers drives continuous condensation and growth of the cold gas phase, as the mixed gas cools and adds mass to the cold reservoir.

Periodic \textit{turbulent box simulation} is another idealized setup that has been studied extensively. 
\rits{When large overdensities (cold gas) are} subjected to a purely turbulent flow, they can be destroyed through continuous fragmentation if radiative cooling is insufficient. \rits{However,} when cooling is strong, the cold \rits{gas can grow in mass} 
(\citealt{Gronke2022, Das2024MNRAS}). 
Thus, the evolution is qualitatively similar to radiative cloud crushing. A similar model has been simulated in the context of multiphase condensation in the ICM, starting from only the hot ICM. These simulations are maintained in rough global thermal balance (unlike the cloud crushing or mixing layer simulations, where the system is cooling overall), motivated by the absence of cluster cooling flows and the presence of AGN bubbles with mechanical power comparable to cooling losses (\citealt{Banerjee2014,Mohapatra2019, Mohapatra2022MNRAS}). In these simulations, turbulence plays a dual role: it can both enhance (by producing density fluctuations) and suppress (by mixing gas before it can cool) multiphase condensation. 
\rits{
These simulations highlight that the thermodynamic state of the medium, and not just local shear or turbulence at the interface, critically governs the onset and persistence of the cold phase in simulations.
For global thermal balance in periodic boxes, in the long-term steady state 
there is no net growth of cold gas mass \citep[$\dot{m}\sim 0$; see fig. 2 of][]{Mohapatra2023}. 
By contrast, in setups such as cloud-crushing and radiative mixing layers, the overall radiative cooling of the system drives a persistent transfer of mass from the hot, diffuse phase to the cold, dense component.}
This role of net cooling or heating in evaporation/condensation is also well studied in thermal instability models applied to interstellar (e.g., see figs. 1-3 of \citealt{Kim2013}) and intracluster medium (e.g., fig. 13 of \citealt{Sharma2010}).

While driven turbulent boxes and radiative cloud-crushing or boundary layer simulations have been studied \rits{extensively, they have largely been explored in isolation. 
However, since turbulence is expected to be ubiquitous in the CGM and galactic outflows, it is essential to examine 
the role of turbulence 
in radiative cloud crushing.}
Galactic scale (\citealt{Vijayan2018, Fielding2018,Steinwandel_2024}) and ISM patch simulations (\citealt{Walch2015,Kim2018ApJ, Tan2023}) reveal \rits{the ubiquity of turbulence in galactic outflows.} 
Observations also hint at the presence of turbulence in different phases of the CGM. The COS-Halo survey suggests turbulent velocities of $\sim 70\ \rm km \ s^{-1}$ in OVI absorption line-widths (\citealt{Werk2013ApJS..204...17W, Faerman_2017}).
Evidence of subsonic turbulence in the cool phase of CGM has also been reported in the CUBS survey (\citealt{ChenQu_2023, Rudie_2019}). 
Additionally, turbulence can be driven at various scales in galaxies and clusters by AGN feedback, mergers, the accretion of pristine gas from the IGM, and the cosmological infall of satellites into the host halo (\citealt{YuanLi2020ApJ, Luo2023}). The simulated tails of ram pressure stripped galaxies (e.g., \citealt{Tonnesen2021ApJ...911...68T}) \rits{tend to be }
\rits{significantly} longer than the \rits{typically} observed stripped tails, \rits{suggesting that their} 
growth, \rits{in reality,} may be limited by background ICM turbulence (\rits{see \citealt{ChenSun2020} for} a sample of stripped galaxy tails in the Coma cluster). 

Although cosmological galaxy formation simulations, particularly those focusing on the CGM, have rapidly improved their spatial/mass resolution (IllustrisTNG \citealt{Nelson2020}; EAGLE \citealt{Oppenheimer2018MNRAS}; SIMBA \citealt{DaveSIMBA2019}; GIBLE \citealt{Ramesh2024MNRAS}), their resolutions are still insufficient to accurately capture the interaction between cold and hot phases. Cool clumps observed in the COS-LRG survey \citep{Zahedy} are typically around \rits{$10-100\ \rm pc$} in size (cf. the right panel of fig. 9 in their paper). The recent JWST/NIRCam observations \rits{of the starburst galaxy M82,} \citep{Fisher2025} reveal 
polycyclic aromatic hydrocarbon (PAH) emission \rits{features (tracing an even colder phase $\lesssim 10^3$ K) with widths as small as $\sim 10 \rm \ pc$}, 
a resolution difficult to achieve in cosmological simulations. In this context, idealized galactic-scale (\citealt{Vijayan2018,Fielding2018, Schneider_2020}) and small-scale simulations (\citealt{GronkeOh2018MNRAS.480L.111G, Armillotta2017, Ji2019, Sparre2020MNRAS, Fielding2020ApJ, FarberGronke2021}) have significantly advanced our understanding of the interplay between radiative cooling and hydrodynamic mixing, processes crucial for cloud survival. In this paper, we continue these explorations, by including driven turbulence on top of radiative cloud-crushing.

\rits{While turbulence is widely known to enhance mixing, and 
in steady state heat the system, its effects on the survival and entrainment of cold gas in galactic winds remain unclear. In the context of radiative cloud-crushing -- where a cold, dense cloud is accelerated and mixed by a hot wind -- one might naturally expect that introducing turbulence would disrupt the cloud more rapidly, enhancing its evaporation and hindering its entrainment. However, this expectation has not been systematically tested. In this work, we revisit the classic cloud crushing problem by explicitly introducing externally driven turbulence into the ambient hot medium, to explore how turbulence modifies the evolution of cold gas in a radiative, multiphase flow.}

\rits{This paper is organized as follows.}
In Section \ref{sec:simulations} we present our simulation setup. In Section \ref{sec:results} we discuss our results, \rits{focusing on the growth and survival of clouds in} 
\textit{cloud crushing simulations with externally driven turbulence}, 
implications of the increasing 
turbulent surface area on the enhancement of cold mass 
\rits{and the dynamics of cloud entrainment.
In Section \ref{sec:discussion}, we comment on the morphological differences induced by turbulence, and discuss the broader implications for simulations and models of galactic outflows. We also explore links to absorption-line diagnostics, particularly modifications to MgII line profiles and an increased projected area of cold gas, before we conclude in Section \ref{sec:conclusions}.
}

\section{Simulation setup}
\label{sec:simulations}
Our simulations are carried out using a modified version of \texttt{PLUTO} (v4.4p2), a conservative hydrodynamic code with a static grid developed by \citet{PLUTO}. For radiative cooling, we use a \texttt{Cloudy} (\citealt{Ferland2017}) generated cooling curve with solar metallicity, \rits{assuming photo+collisional ionization equilibrium for the plasma in the presence of} 
\citet{HaardtMadau2012ApJ} extragalactic UV radiation at redshift 0. 
A temperature floor at $4\times 10^4\rm \ K$ is assumed in all the simulations to mimic the effect of photoheating. We evolve the Euler equations in \texttt{PLUTO}, with turbulent forcing added as a source term. The ideal gas equation of state relates the density $\rho$, pressure $P$, and temperature $T$. The following equations are evolved in our simulations,
\begin{subequations}
\begin{align}
    &\frac{\partial \rho}{\partial t} + \nabla .(\rho {\bf v}) = 0,\\
    &\frac{\partial (\rho {\bf v})}{\partial t} + \nabla .(\rho {\bf v} \otimes {\bf v}+P\mathbb{I}) =  {\bf F}, \\
    &\frac{\partial e}{\partial t} + \nabla .((e+P){\bf v}) = {\bf F}.{\bf v} - \mathcal{L}(n, T),
\end{align}
\end{subequations}
where $e=\rho v^2/2 + P/(\gamma -1)$ is the total energy density, $\rho$ is the mass density, ${\bf v}$ is the fluid velocity, $P=\rho k_BT/(\mu m_p)$ is the pressure, $T$ is the temperature, $\mu$ is the mean molecular mass, $m_p$ is the proton mass, $k_B$ is the Boltzmann constant, $\mathcal{L}(n,T)$ is the cooling rate per unit volume and $\gamma=5/3$ is the adiabatic index. The term ${\bf F}$ is the applied turbulent force per unit volume, which follows the stochastic Ornstein-Uhlenbeck process (\citealt{ESWARAN1988257}; as elaborated in \citealt{Mohapatra2019}), the details of which are further explained in Section \ref{subsec:turb_forcing}. Here, the cooling rate per unit volume is,
\begin{equation}
    \mathcal{L}(n, T)=n_H^2\Lambda(T)\; ,
\end{equation} 
where $\Lambda(T)$ is the temperature-dependent cooling function and \rits{$n_H= \rho X_H/m_p$} is the \rits{total} hydrogen number density \rits{and $X_H$ is the total hydrogen mass fraction}.

\par We carry out cloud-crushing simulations to study the effect of external turbulent forcing on cloud survival in galactic winds. The details of these simulations are outlined in Section \ref{subsec:cc_simulations}, and Table \ref{tab:cloud_crushing_turb} lists the various parameters of the cloud-crushing simulations performed. In Section \ref{subsec:turb_forcing}, we elaborate on the forcing algorithm.

\begin{table*}
\centering
\setlength{\tabcolsep}{9pt} 
\renewcommand{\arraystretch}{0.99} 
    \caption{Overview of various parameters considered in our cloud crushing simulations with and without externally driven turbulence.${}^{\dag}$ 
    }
    \vspace{-3pt} 
    \noindent\hspace{1.1em}\makebox[0.84\linewidth]{\rule{\dimexpr0.84\linewidth}{0.6pt}}
    \vspace{-1pt}
    \begin{tabular}{cccc}
        \hline
        $\mathcal{M}_{\rm wind}$${}^{{\color{blue}a}}$ & $t_{\rm cool,mix}/t_{\rm cc}$ & $\mathcal{M}_{\rm turb}$${}^{{\color{blue}b}}$ & Cloud growth ${}^{{\color{blue}c}}$ \\
        \hline
        1.5 & 0.001 & 0    &  \checkmark \\
         -  & 0.001 & 0.22 &  \checkmark \\
         -  & 0.001 & 0.30 &  \checkmark \\
         \smallskip
         -  & 0.001 & 0.67 &  \checkmark \\ 
         -  & 0.01  &  0   &  \checkmark \\
         -  & 0.01  & 0.22 &  \checkmark \\
         -  & 0.01  & 0.30 &  \checkmark \\
        \smallskip
         -  & 0.01  & 0.67 &  \checkmark \\
         -  & 0.1   &  0   &  \checkmark \\
         -  & 0.1  &  0.20 &  \checkmark \\
         -  & 0.1  &  0.30 &  \checkmark \\
         \smallskip
         -  & 0.1  &  0.67 &  \checkmark \\
         -  & 0.3   &  0   &  \checkmark \\
         -  & 0.3  &  0.18 &  \checkmark \\ 
         -  & 0.3  &  0.30 &  \checkmark \\
         -  & 0.3  &  0.54 &  \checkmark \\
         \smallskip
         -  & 0.3  &  0.67 &  \checkmark \\ 
         -  & 1.0  &   0   &  \checkmark \\
         -  & 1.0  &  0.15 &  \checkmark \\ 
         -  & 1.0  &  0.30 &  \checkmark \\ 
         -  & 1.0  &  0.50 &  $?$ \\ 
         \smallskip
         -  & 1.0  &  0.67 &  \tikzxmark \\
            &      &       &             \\
        \end{tabular}
        \hspace{-0.5pt}    
        \vrule width 0.8pt 
        \vrule width 0.8pt 
        \begin{tabular}{cccc}
        \hline
        $\mathcal{M}_{\rm wind}$ & $t_{\rm cool,mix}/t_{\rm cc}$ & $\mathcal{M}_{\rm turb}$ & Cloud growth\\
        \hline
        1.5 &  4.0 &  0    & \tikzxmark \\
        -   &  4.0 &  0.29 & \tikzxmark \\
        -   &  4.0 &  0.54 & \tikzxmark \\
        \smallskip
        -   &  4.0 &  0.67 & \tikzxmark \\
        -   &  6.0 &  0.12 & \tikzxmark \\
        \smallskip
        -   &  6.0 &  0.60 & \tikzxmark \\
        \smallskip
        -   & 10.0 &   0   & \tikzxmark \\
        \smallskip
        0.65 & 0.03 & 0.67 & \checkmark \\
        \smallskip
        -   &  0.3 &  0.48 & \tikzxmark \\
        \smallskip
        0.56 & 0.3 &  0.67 & \tikzxmark \\
        0.50 & 0.1 &  0.29 & \checkmark \\
        \smallskip
         -   & 0.1 &  0.67 & \checkmark \\
         -   & 0.3 &  0.45 & \checkmark \\
         -   & 0.5 &  0.45 & \tikzxmark \\
        \smallskip
         -   & 0.5 &  0.67 & \tikzxmark \\
         -   & 1.0 &  0.29 & \tikzxmark \\
         -   & 1.0 &  0.45 & \tikzxmark \\
        \smallskip
         -   & 1.0 &  0.67 & \tikzxmark \\
         \smallskip
         -   & 1.5 &    0  & \checkmark \\
         \smallskip
         -   & 3.0 &    0  & \checkmark \\
         -   & 6.0 &    0  & \tikzxmark \\
    \end{tabular}
    \vspace{1.8pt}
    \noindent\makebox[0.9\linewidth]{\rule{\dimexpr0.85\linewidth}{0.7pt}}
    \begin{tablenotes}
        \item ${}^{\dag}$ {\fontsize{7pt}{7.5pt}\selectfont All simulations are carried out with a cloud of overdensity $\chi=100$ relative to the wind, which is at temperature $T_{\rm wind}=4\times 10^6$ K and $P/k_{\rm B}=2\times10^4\ \rm cm^{-3}\ K$.}
        \item ${}^{{\color{blue}a}}$ {\fontsize{7pt}{7.5pt}\selectfont The wind Mach number $\mathcal{M}_{\rm wind}$, is defined as the ratio of the initial relative velocity between the cloud and the wind $v_{\rm wind}$ to the sound speed in the hot medium $c_{\rm s,wind}$. Values identical to those in the preceding row are indicated by "-".}
        \item ${}^{{\color{blue}b}}$ {\fontsize{7pt}{7.5pt}\selectfont $\mathcal{M}_{\rm turb}=v_{\rm turb}/c_{\rm s, wind}$ is the ratio of the (rms) velocity in the domain (at driving scale) to the sound speed in the hot phase. Entries labeled “0” correspond to runs with no turbulence.}
        \item ${}^{{\color{blue}c}}$ {\fontsize{7pt}{7.5pt}\selectfont Indicates whether the cloud survives (\checkmark) or is destroyed (\tikzxmark), with destruction defined as the cold gas mass (gas below $3\times T_{\rm cl}$) dropping below 10\% of the initial cloud mass.  Runs that exhibit a clear negative slope in mass evolution by the end of the simulation, but retain more than 10\% of the cold gas, are marked with "$?$".
        }
    \end{tablenotes}
    \label{tab:cloud_crushing_turb}
\end{table*}

\subsection{Simulation details and parameters}\label{subsec:cc_simulations}
Our simulation framework consists of a spherical cold ($\sim 4\times 10^4\rm \ K$) cloud with radius $R_{\rm cl}$ moving through a hot, turbulent ambient medium.
For this, we first initialize a uniform hot phase in a periodic domain, characterized by a temperature of $4\times 10^6\rm \ K$, and pressure $P/k_{\rm B}=2\times 10^4\rm \ cm^{-3}\ K$. The domain is continuously stirred with solenoidal driving at large scale $L_{\rm eddy}=40R_{\rm cl}$ for over two eddy turnover timescales (see Section \ref{subsec:turb_forcing} for details). This yields a root mean square (rms) turbulent velocity $v_{\rm turb} \propto L_{\rm eddy}/t_{\rm eddy}$ in the hot phase such that the rms turbulent Mach number is $\mathcal{M}_{\rm turb}=v_{\rm turb}/c_{\rm s, wind}$, where $c_{\rm s,wind}$ represents the sound speed in the hot medium.
The cloud is introduced isobarically into this turbulent medium with a density contrast $\chi$, reflecting a realistic scenario in which a cloud driven out of the galactic disk, encounters a wind \rits{rendered} 
turbulent \rits{by} 
multiple supernovae bursts. 
We impose an initial velocity on the cloud to establish a relative motion with speed $v_{\rm wind}$ corresponding to a Mach number $\mathcal{M}_{\rm wind}=v_{\rm wind}/c_{\rm s, wind}$ \rits{between the cloud and the wind}. All simulations are performed in a three-dimensional Cartesian domain with the initial cloud motion directed along the $\hat{x}$-axis. The boundary conditions remain periodic in all directions and turbulence is constantly driven in the box \rits{(in the hot phase; see further discussion in \S \ref{subsec:turb_forcing})}. 

\rits{In all our simulations, the grid size is chosen such that the initial cloud is resolved by 8 grid cells across it, i.e., $R_{\rm cl}/d_{\rm cell}=8$}.
We sample a range of subsonic turbulent Mach numbers $\mathcal{M}_{\rm turb}$ (ranging from 0.1, 0.7) motivated by observations and because supersonic turbulence would not be sustainable \rits{due to turbulent heating} and would quickly transition to becoming subsonic. 
A domain size of ($9L_{\rm eddy}, L_{\rm eddy}, L_{\rm eddy}$) is used for most of the simulations, with an overall resolution of $(2880\times 320\times 320)$. For $\mathcal{M}_{\rm turb}>0.5$, we use a shorter length along the $\hat{x}$-direction and a larger orthogonal extent such that the size of the box is ($3L_{\rm eddy}, 2L_{\rm eddy}, 2L_{\rm eddy}$), with a 
resolution of ($960\times 640\times 640$). This helps us to reduce the computational cost to capture the orthogonal spread of the cloud
\footnote{Note that this does not affect our results as we would later see in Section \ref{subsec:entrainment}, the clouds are easily entrained at high Mach numbers and this longitudinal extent of $3L_{\rm eddy}$ is sufficient to capture the head and tail of the cloud.}, especially for short $t_{\rm cool,mix}/t_{\rm cc}$. 

\par We use the HLLC (Harten, Lax, van Leer Contact) solver, with RK3 time-stepping and a linear reconstruction scheme. The initial cloud is marked with a passive scalar $C$ (\rits{initialized} to 1 inside the cloud and 0 outside) and the background wind is prevented from cooling by \rits{setting the radiative cooling rate to zero} 
for any grid cell with tracer value $C<10^{-4}$, \rits{approximating the effect of additional heating, which maintains the wind as a quasi-static hot reservoir.}

For our parameter study, we focus on an overdensity of $\chi = 100$ and a \rits{supersonic wind with Mach number} $\mathcal{M}_{\rm wind}\equiv v_{\rm wind}/c_{\rm s,wind}=1.5$. 
We varied the efficiency of cooling and the magnitude of the external turbulence. In particular, we varied the ratio $t_{\rm cool,mix}/t_{\rm cc}$ from $10^{-3}$ to $10^{1}$ where $t_{\rm cc}=\chi^{1/2} R_{\rm cl}/v_{\rm wind}$ is the classical `cloud crushing' time \citep{Klein1994ApJ,Scannapieco_2015} and $t_{\rm cool,mix}$ is the cooling time of the `mixed' medium defined as being at \rits{a temperature and density corresponding to} the geometric mean between the wind and cloud temperatures and densities \citep{BegelmanFabian1990MNRAS.244P..26B}. Earlier radiative cloud-crushing simulations \rits{with a laminar wind} found that cold clouds survive and grow if $t_{\rm cool,mix}/t_{\rm cc} < 1$ (\citealp{GronkeOh2018MNRAS.480L.111G} but see discussion, e.g., in \citealp{LiHopkins2020,Sparre2020MNRAS,Kanjilal2021}). Here we study how turbulence affects the growth and destruction of cold gas depending on the strength of cooling and turbulence -- 
which we will elaborate upon in Section \ref{subsec:growth_cc}. 

\subsection{Turbulent forcing method}
\label{subsec:turb_forcing}
We adopt the turbulent forcing scheme described in \citet{Mohapatra2019}. Briefly, \rits{our turbulent forcing algorithm is} a spectral forcing method \rits{that employs} the stochastic Ornstein-Uhlenbeck process to model a continuous turbulent acceleration ${\bf a}$ (\citealp{ESWARAN1988257, Schmidt2006, Federrath2010}). \rits{The resulting turbulent force ${\bf F}$ is computed at each timestep in real space, such that at position $x$, the force at the $n$-th timestep is given by,}
\footnote{Eq. 8 in \citet{Mohapatra2019} has a typo and should not have the integral over space.}
\begin{equation}
    {\bf F}^n({\bf x})= \rho({\bf x})\; \text{Re} 
    \left(
    \sum_{|k|=k_{\rm min}}^{k_{\rm max}}{\bf a}_{\bf k}^n
    \; e^{-i{\bf k}.{\bf x}}\;  \right) 
\end{equation}
where, $k_{\rm min}$ and $k_{\rm max}$ are the modes to which the forcing is limited. The term ${\bf a}_k^n$ is the acceleration in the $n$-th timestep in Fourier space \rits{of the discretized simulation domain}. Its time evolution follows the stochastic Ornstein–Uhlenbeck process:
\begin{equation}
    {\bf a}_k^n = f{\bf a}_k^{n-1} + \sqrt{1-f^2}{\bf a'}_k^{n},
\end{equation}
where $f$ is an exponential decay factor $f=\exp(-\delta t^n/\tau_c)$ for n>0 (0, if $n=0$), $\delta t^n$ is the $n$-th timestep size, and $\tau_c$ is the auto-correlation timescale that controls the temporal coherence of the forcing.  
For our simulations, $\tau_c=t_{\rm eddy}$ throughout. \rits{The 
acceleration ${\bf a'}_k^n$ at the $n$-th timestep in the Fourier space is sampled from a Gaussian random number generator with amplitude $A_{\rm turb}$, such that ${\bf a'}_k^n = A_{\rm turb} \mathcal{N}(0,1)$.}
The driving is purely solenoidal as we subtract the component \rits{of acceleration} along ${\bf k}$ in the Fourier space, such that
\[{\bf a}_k^n\ \rits{\leftarrow}\ {\bf a}_{\bf k}^n-({\bf a}_{\bf k}^n.{\bf k}){\bf k}/|{\bf k}^2|.\]
We drive only at large scales such that  $K_{\rm driving}\in (0,\sqrt{2})$, where $K_{\rm driving}=k/(2\pi)$, and $k = 2\pi / L_{\rm eddy}$ ($L_{\rm eddy}$ is the driving scale, which is the same as the extent of the simulation domain perpendicular to the wind, unless otherwise stated). 

Since turbulence is forced throughout the computational domain, both the hot and cold phases would experience acceleration due to turbulent forcing. To prevent any unphysical acceleration of the dense cloud structures, we implement a density-weighted forcing scheme where the external force for cold gas (cells with temperature $T<3 T_{\rm cl}$) is scaled inversely with the density ($F\propto 1/\rho$)\footnote{As the typical density contrast between the cloud and wind is $\sim 100$, this ansatz ensures that the cloud does not get accelerated by turbulent forcing}.
Additionally, we ensure that no net momentum is introduced in the simulation domain due to external forcing. This is achieved by calculating the \rits{(volume)} average momentum in all directions $\langle\rho \delta{\bf v}({\bf x})\rangle$ (where $\delta{\bf v}= {\bf F}({\bf x})/\rho({\bf x})\rits{\delta t}$ is the additional velocity introduced by forcing) and subtracting it out (such that $\langle\rho \delta{\bf v}({\bf x})\rangle$=0 due to external forcing).

\section{Results}
\label{sec:results}

\begin{figure*}
    \includegraphics[width=\textwidth]{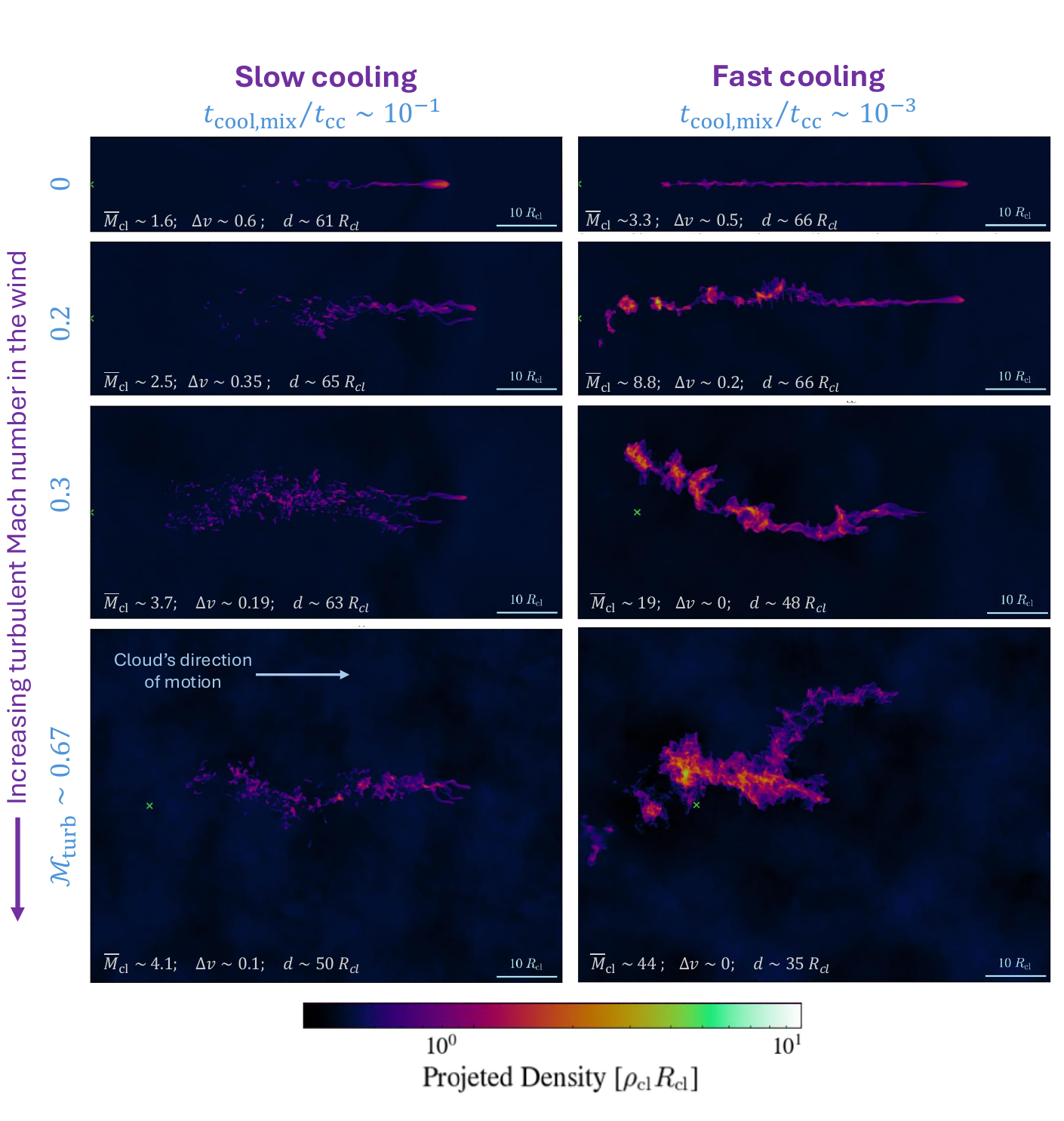}
     \caption{
     Projected column density of a cloud moving through a turbulent wind at various strengths of turbulent Mach number $\mathcal{M}_{\rm turb}$ (increasing from \emph{top to bottom}). The [\emph{left column}] shows the cloud evolution in the weak cooling regime ($t_{\rm cool,mix}/t_{\rm cc}=10^{-1}$), while the [\emph{right column}] shows the state in the strong cooling regime ($t_{\rm cool,mix}/t_{\rm cc}=10^{-3}$). All the snapshots are taken at $8 t_{\rm cc}$, where $t_{\rm cc}\sim \chi^{1/2}R_{\rm cl}/v_{\rm rel}$ is the standard cloud crushing time and the cloud is moving towards right (along $\hat{x}$; as indicated by the arrow in bottom left panel) with velocity $v_{\rm rel}$ in a hot medium that is continuously stirred by external turbulent forcing. \rits{All panels show selected regions of the simulation domain, with a cyan cross marking the cloud’s initial position whenever visible. }
     The turbulent Mach number in the hot medium 
     $\mathcal{M}_{\rm turb}=v_{\rm turb} /c_{\rm s, wind}$ ($v_{\rm turb}$ and $c_{\rm s, wind}$ are the rms turbulent velocity and sound speed in the hot medium) is indicated on the left.
     Note that the morphology of tails of cold gas with and without turbulence is very different. 
     Clouds in a turbulent wind are much more \rits{fluffier} 
     and stretched in the orthogonal direction, in comparison to the elongated streaks of cold mass when the cloud faces a uniform wind (top panels).
     Increased turbulent forcing enhances the cold gas mass (see $\overline{M}_{\rm cl}=M_{\rm cl}/M_{\rm cl,0}$ reported in each panel). 
     The small relative velocity $\Delta v\sim |v_{\rm cl}-v_{\rm wind}|/v_{\rm rel}$ and displacement of the cloud $d$ observed at high $\mathcal{M}_{\rm turb}$, demonstrates that the cloud grows via 
     a continuous accretion of mass and momentum 
     from the increasing turbulent mixing layers in presence of stronger and stronger turbulent forcing.  
     A curated playlist of videos  illustrating the evolution of cold gas are available here: \url{https://www.youtube.com/playlist?list=PLuwSozndVCNJzjKWqRO8u-GTCE0U8Zny2} .
     }
    \label{fig:snapshotsProjected}
\end{figure*}

\begin{figure*}
    \centering
     \includegraphics[width=\textwidth]{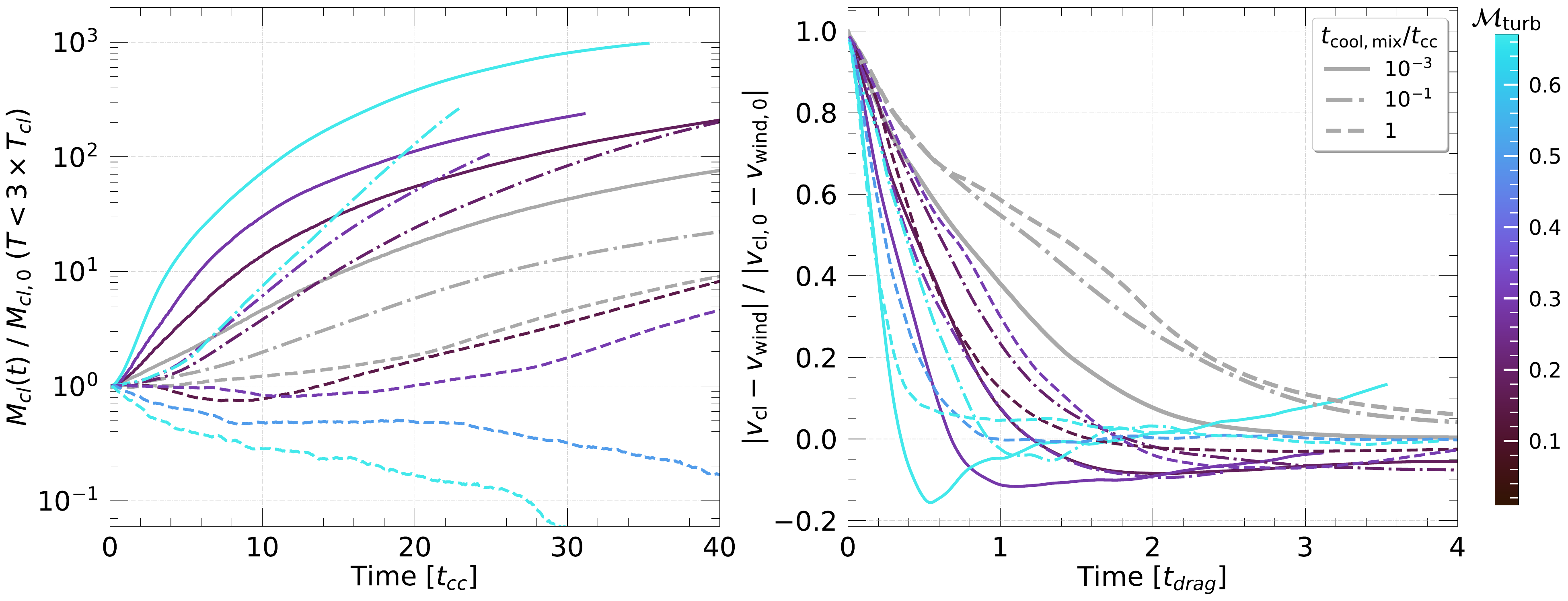}
    \caption{
    [\textit{Left panel}]:
    Evolution of the cold mass $M_{\rm cl}$ (normalized by the initial cloud mass $M_{\rm cl,0}$) in units of cloud crushing time $t_{\rm cc}$ at various turbulent Mach numbers $\mathcal{M}_{\rm turb}=v_{\rm turb}/c_{\rm s, wind}$ and the ratio of the cooling time of the mixed gas to the cloud crushing time $t_{\rm cool, mix}/t_{\rm cc}$. 
    [\textit{Right panel}]: The relative velocity between the cloud and wind along the x-direction $v_{\rm cl}-v_{\rm wind}$ in units of the initial relative velocity $|v_{\rm cl,0}-v_{\rm wind,0}|$ plotted as a function of drag time $t_{\rm drag}\sim \chi R_{\rm cl}/v_{\rm wind}$. 
    In both panels, the gray lines represent evolution in a laminar wind while the colored lines mark the evolution in a turbulent wind with strengths as indicated in the colorbar. In the left panel, the solid lines show how cold mass growth is enhanced in the fast cooling regime ($t_{\rm cool, mix}/t_{\rm cc} \sim 10^{-3}$). With increasing turbulence (higher $\mathcal{M}_{\rm turb}$ in \textit{magenta} and \textit{cyan} lines). Conversely, when cooling is weak ($t_{\rm cool, mix}/t_{\rm cc}\sim 1$),
    increasing turbulence hinders mass growth. This is demonstrated by the \textit{dashed colored} lines, which show a reduced cold mass growth compared to the \textit{dashed gray} line (without turbulence).
    As the turbulent forcing increases, it stretches and mixes 
    the cold gas, 
    leading the cloud to transition into the destruction regime (\textit{dashed cyan} line ). Interestingly, in the right panel, the colored lines (with turbulence) 
    fall steeply compared to the gray lines (without turbulence), suggesting a faster entrainment of clouds in winds with continuous turbulent forcing.
    }
    \label{fig:mass_growth}
\end{figure*}
In this section, we present results from our simulations of cloud crushing with continuous turbulent forcing. In Section \ref{subsec:growth_cc}, we discuss the implications of turbulent driving at different amplitudes on the growth and survival of clouds in an otherwise uniform wind. We elucidate 
that clouds in a turbulent wind form shorter and clumpier structures rather than the elongated tails formed in a laminar wind. In Section \ref{subsec:growth_rate_and_area}, we analyze the interplay 
between the mass growth rate and an increasing area of the mixing layers between the cloud and the wind.
Section \ref{subsec:entrainment} presents our finding about the faster entrainment of clouds in turbulent winds.

\subsection{Growth of clouds in a turbulent galactic wind}\label{subsec:growth_cc}
We first discuss the implications 
of continuous turbulent forcing on the growth and survival of clouds in an otherwise uniform wind. As discussed in Section \ref{subsec:cc_simulations} and outlined in Table \ref{tab:cloud_crushing_turb}, we carry out simulations with different strengths of turbulent Mach numbers $\mathcal{M}_{\rm turb}=v_{\rm turb}/c_{\rm s, wind}$.

Fig. \ref{fig:snapshotsProjected} shows the projected number densities 
along the $\hat{z}$-direction (orthogonal to the direction of motion of the cloud $\hat{x}$) for different growth regimes (depending on $t_{\rm cool, mix}/t_{\rm cc}$) in the two columns. From top to bottom, the strength of turbulent forcing (\rits{as measured by} 
$\mathcal{M}_{\rm turb}$) increases from $0$ to $0.67$. All the panels show the state of the cloud at $8t_{\rm cc}$, where $t_{\rm cc}\sim \chi^{1/2}R_{\rm cl}/v_{\rm rel}$ is the standard cloud crushing time for a cloud of size $R_{\rm cl}$ moving with a velocity $v_{\rm rel}$ in the wind's frame of reference \citep{Klein1994ApJ}.
The right column in Fig. \ref{fig:snapshotsProjected}, shows the evolution of a cloud in the fast cooling regime ($t_{\rm cool, mix}/t_{\rm cc}=10^{-3}$) with different turbulent driving strengths - without turbulence in the top right and with turbulence of $\mathcal{M}_{\rm turb}\sim [0.2, 0.3, 0.67]$ in the second, third and fourth rows, respectively. It is interesting to note the change in the morphology of the dense structures with and without turbulence. The otherwise elongated tail in a laminar wind (top right panel) grows to form a \rits{fluffy} 
and stretched blob of cold gas with regions of high density gas as the turbulent Mach number $\mathcal{M}_{\rm turb}$ in the wind increases. On the other hand, if the cooling is slow ($t_{\rm cool, mix}/t_{\rm cc}=0.1$), the left-hand panel of Fig. \ref{fig:snapshotsProjected} suggests that turbulence enhancement of cloud progresses slowly. In all cases where the cloud grows, we find that turbulence limits the elongation of the cold dense structure down the wind and produces a shattered and orthogonally spread-out structure of dense gas (discussed further in \S~\ref{subsec:morphology_with_turb}). The head-tail features are also strikingly different \rits{across different turbulent Mach numbers}. 
While the case \rits{without} turbulence shows a clear head with a linear 
streak in the tail; the clouds in a turbulent wind have low column densities in the head and high columns in the tail. This is because, the regions in the tail are most likely to become co-moving with the wind, the shear reduces and this results in an enhancement of cold gas with continuous accretion in those regions if cooling can progress fast.

In each panel of Fig. \ref{fig:snapshotsProjected}, we report the cloud mass in terms of its initial mass $\overline{M}_{\rm cl}= M_{\rm cl}(t)/ M_{\rm cl,0}$, the relative velocity (along $x$) between the cloud and the wind in units of the initial velocity difference, $\Delta v=|v_{\rm cl}-v_{\rm wind}|/v_{\rm rel, 0}$ as well as the distance traveled by the cloud $d$ which is measured as the 90th percentile of a PDF of $x-$ coordinates of cold gas in the domain 
(cells at temperature $T<3 T_{\rm cl}$). In both panels, the clouds facing turbulent wind are more spread out in the orthogonal direction compared to the laminar case. When cooling is slow (left panel; $t_{\rm cool, mix}/t_{\rm cc}\sim 0.1$), this spread is smaller and the mass is enhanced by a factor of a few. But the relative velocity between the cloud and the wind reduces quickly. 
In the fast cooling regime (right column; $t_{\rm cool, mix}/t_{\rm cc}=10^{-3}$), entrainment is even faster, the cloud spreads over a larger volume and the mass grows by $\sim$1dex at $\mathcal{M}_{\rm turb}$ of 0.3 and 0.67. This is \rits{an interesting behavior} and we will return to this point in 
Section \ref{subsec:entrainment}.

So far, the density projections in Fig. \ref{fig:snapshotsProjected} show a clear trend of cold gas mass enhancement with increasing turbulence and highlight \rits{a gradual change in the} morphological features of clouds moving in \rits{an increasingly} turbulent wind.
We now turn our attention to \rits{a quantitative} 
analysis of mass growth in the presence of continuous turbulent driving in an otherwise laminar wind. 

The left panel of Fig. \ref{fig:mass_growth} shows the evolution (in units of cloud crushing time $t_{\rm cc}$) of cold mass (all gas at temperature $T<3\times T_{\rm cl}$) normalized by the initial cloud mass, for different cooling regimes (\rits{$t_{\rm cool,mix}/t_{\rm cc} =  [10^{-3},\ 0.1,\ 1] \iff [\rm{solid,\ dash-dotted,\ dashed}]$ lines} 
) at several strengths of turbulent forcing. The gray lines mark the evolution for a laminar wind ($\mathcal{M}_{\rm turb}=0$; no turbulence) while the colored lines are color-coded according to the rms turbulent Mach number $\mathcal{M}_{\rm turb}$ as indicated by the colorbar. Increasing $\mathcal{M}_{\rm turb}$ results in \rits{an increase in} cold mass growth by an order of magnitude in the fast cooling regime (compare the solid blue curve at $\mathcal{M}_{\rm turb}=0.67$ with the solid gray line at $\mathcal{M}_{\rm turb}=0$). 
For a slow cooling regime (dashed lines), turbulence suppresses the cold mass as $\mathcal{M}_{\rm turb}$ increases. When cooling is slow, shear 
due to turbulence can be high enough to change the fate of the cloud from growing at $\mathcal{M}_{\rm turb}=0.2$ (dashed line in magenta) to being destroyed at $\mathcal{M}_{\rm turb}=0.67$ (dashed line in cyan). \rits{Overall, the growing clouds grow faster and the destroyed ones are destroyed faster with turbulence.}
We note that the hot gas temperature increases by a factor of 4-10 of the initial value at later times, as work done by
turbulent forcing heats 
the wind thereby increasing the sound speed $c_s$. We have checked that our results are unaffected by this increase in temperature by additional simulations with a temperature ceiling above $T_{\rm hot}$ (not shown here for clarity). 

The right panel of Fig. \ref{fig:mass_growth} shows the evolution of the velocity difference along $x$ between the cloud (gas at temperature $T<3\times T_{\rm cl}$) and the hot ambient medium $v_{\rm cl}-v_{\rm wind}$, normalized with the initial velocity difference $|v_{\rm cl,0}-v_{\rm wind,0}|$. The time is expressed in units of the standard cloud crushing time $t_{\rm cc}$.
The line color and styles are \rits{sampled from the attached colorbar of turbulent (rms) Mach number $\mathcal{M}_{\rm turb}$ in the hot phase (as discussed above for the left panel of Fig. \ref{fig:mass_growth}).} 
similar to those in the left panel.
Increasing $\mathcal{M}_{\rm turb}$ results in faster entrainment, especially for the cyan lines with $\mathcal{M}_{\rm turb}=0.67$ in different linestyle, as they fall steeply compared to the gray lines. We will discuss this in more detail in Section \ref{subsec:entrainment}.

\begin{figure}
    \includegraphics[width=\columnwidth]{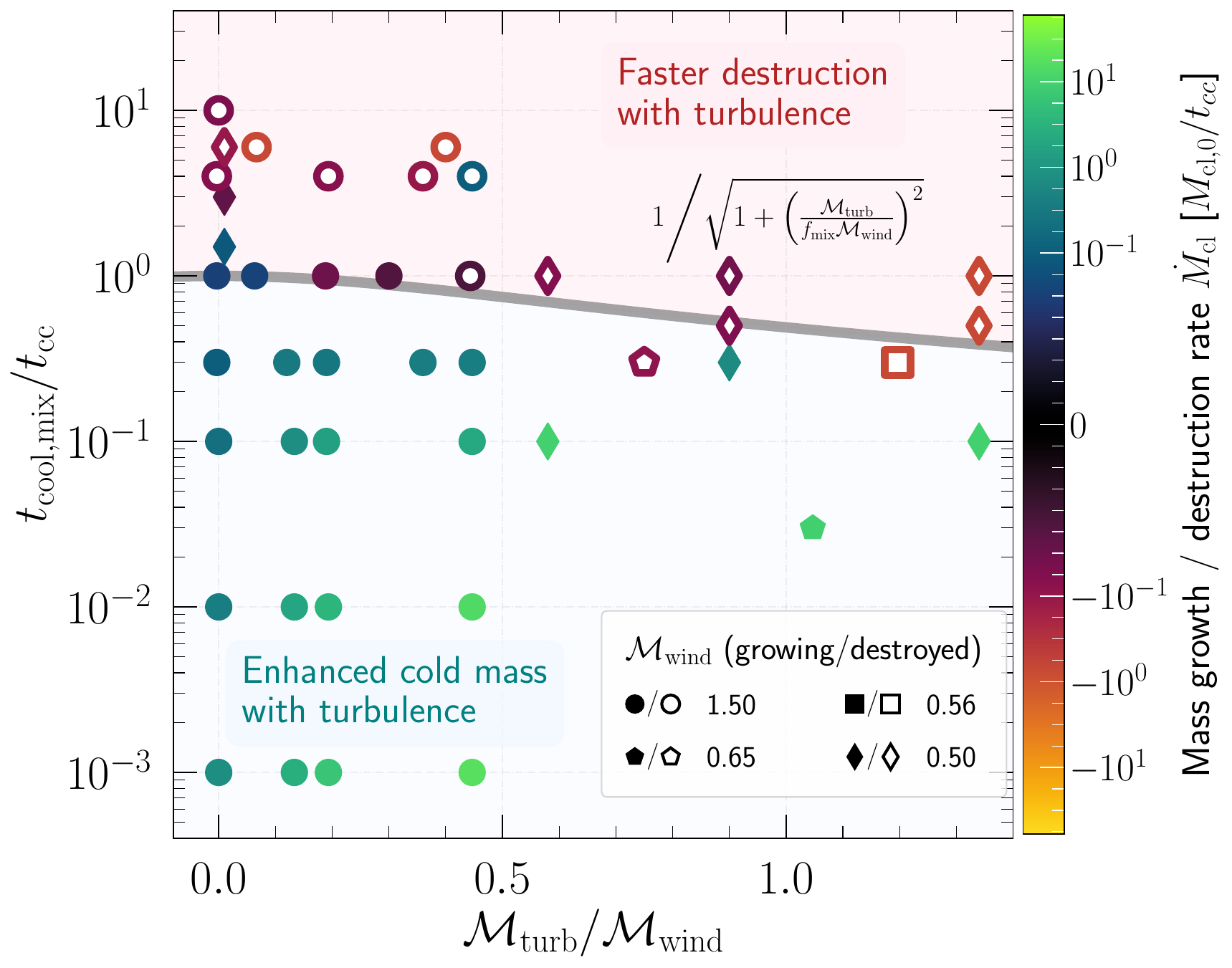}
    \caption{The parameter space defined by the ratio $t_{\rm cool, mix}/t_{\rm cc}$ and the ratio of turbulent to wind Mach numbers $\mathcal{M}_{\rm turb}/\mathcal{M}_{\rm wind}$, illustrating the regimes for growth and destruction of clouds in cloud-crushing simulations with driven turbulence. \rits{Different markers represent various wind Mach numbers: circles, pentagons, squares and diamonds correspond to $\mathcal{M}_{\rm wind}=1.5, 0.65, 0.56,$ and $0.5$, respectively. \textbf{Filled markers} indicate growing clouds, while the \textbf{hollow markers}} denote clouds that are eventually destroyed. We define a cloud as destroyed if the cold gas mass (see Fig. \ref{fig:mass_growth}) falls below 10\% of the initial cloud mass $M_{\rm cl,0}$. Each point in the phase space is color-coded according to the cold gas mass growth rate $\dot{M}_{\rm cl}$ (in units of $M_{\rm cl,0}/t_{\rm cc}$) \rits{measured} at $10\ t_{\rm cc}$, \rits{as indicated by the colormap.
    For clouds that are destroyed, the color reflects their destruction rate; }if destruction occurs before $10\ t_{\rm cc}$, a fixed destruction rate of $\dot{M}_{\rm cl}=-0.7 M_{\rm cl,0}/t_{\rm cc}$ is used instead.
    \rits{As we move from left to right in the plot, the filled circles become progressively brighter green compared to the $\mathcal{M}_{\rm turb}=0$ case, indicating an enhancement of cold gas mass growth, especially at smaller values of $t_{\rm cool, mix}/t_{\rm cc}$. }
    The modified criterion \rits{ $t_{\rm cool, mix}/\tilde{t}_{\rm cc}<1$, where $\left.\tilde{t}_{\rm cc}= t_{\rm cc} \middle/\sqrt{1+\left(\mathcal{M}_{\rm turb}/f_{\rm mix}\mathcal{M}_{\rm wind}\right)^2}\right.$ with $f_{\rm mix}\sim0.6$, is}
    shown in the \textit{solid gray} line, which clearly separates the phase space into growth (shaded region in \textit{blue}) and destruction (shaded region in \textit{pink}) regimes.
    }
    \label{fig:overview_cc_turb}
\end{figure}

Figure~\ref{fig:overview_cc_turb} provides an overview of the parameter space explored in our cloud crushing simulations (cf. Table \ref{tab:cloud_crushing_turb}) with continuous turbulent forcing in the $t_{\rm cool, mix}/t_{\rm cc}$ versus $\mathcal{M}_{\rm turb}/\mathcal{M}_{\rm wind}$ plane.
The \rits{filled symbols in Fig.~\ref{fig:overview_cc_turb} denote cloud survival while the hollow ones} mark the destruction of the cloud, where we refer to the cloud as destroyed whenever the cold gas mass $M_{\rm cl}(t)$ in the simulation domain falls below 10\% the initial value $M_{\rm cl,0}$. The filled and hollow markers are colored according to the rate of mass growth $\dot{M}_{\rm sat}$ at $10t_{\rm cc}$ ($\dot{M}_{\rm sat}$ in units of the ratio of initial cloud mass $M_{\rm cl,0}$ and $t_{\rm cc}$). 
For cases showing destruction, we color code the points by their destruction rate; if destruction occurs before $10\ t_{\rm cc}$, a fixed destruction rate of $\dot{M}_{\rm cl}=-0.7 M_{\rm cl,0}/t_{\rm cc}$ is used instead.

Note that the gray line in Fig.~\ref{fig:overview_cc_turb}, which separates the simulations with surviving and destroyed cold gas, $t_{\rm cool,mix}/t_{\rm cc}=1/\sqrt{1+(\mathcal{M}_{\rm turb}/f_{\rm mix}\mathcal{M_{\rm wind}})^2}$ (with a fudge factor $f_{\rm mix}\approx 0.6$ weighing the relative importance of shear and turbulence) which separates the growing and destruction regime quite well. An exception occurs in cases with no turbulence ($\mathcal{M}_{\rm turb}=0$) and low wind velocity ($\mathcal{M}_{\rm wind}=0.5$) runs, which seem to show a facilitated survival. 
This failure of the classical survival criterion is likely due to the increased time the mixed gas spends to accumulate in the tail and radiatively cool, aided by the slower advection and weaker shear. 
A systematic study of this regime, however, is beyond the scope of this work.

The boundary seen in Fig.~\ref{fig:overview_cc_turb} can be recast to an altered version of the `cloud crushing time', defined as
\begin{equation}
    \tilde t_{\rm cc} \equiv \chi^{1/2} \frac{R_{\rm cl}}{\sqrt{v_{\rm wind}^2+(v_{\rm turb
    }/f_{\rm mix})^2}},
    \label{eq:tcc_tilde}
\end{equation}
i.e., which takes both the turbulent as well as the wind velocity into account.
Note that in Eq.~\eqref{eq:tcc_tilde}, we used $v_{\rm rms}$ as a proxy for the turbulent velocity $v_{\rm turb}$. More precisely, this should be $u'(L_{\rm cold}) \approx v_{\rm turb}(L_{\rm cold}/L_{\rm eddy})^{1/3}$, where $L_{\rm cold}$ is the scale of the cold gas that turbulence is acting upon. While we can constrain this scale to lie within $r_{\rm cl} < L_{\rm cold} < L_{\rm eddy}$, its exact value remains uncertain. However, since the cold gas rapidly stretches to scales of $\sim$ tens of $r_{\rm cl}$ (cf.~\S~\ref{subsec:morphology_with_turb} and Fig.~\ref{fig:tail_length}), and $L_{\rm eddy} = 40\,r_{\rm cl}$ in our fiducial setup, the approximation $u'(L_{\rm cold}) \approx v_{\rm rms}$ is reasonable. 
This uncertainty of the relevant $u'$ as well as the fact that destruction usually occurs at a few $t_{\rm cc}$ \citep[e.g.][]{Scannapieco_2015} are absorbed in the fudge factor $f_{\rm mix}.$


In summary, external turbulence can lead to both facilitated destruction of the cold gas (if $t_{\rm cool,mix}/t_{\rm cc}$ is close to the survival threshold) and enhanced growth (in the faster cooling regime). However, the regime in which cold gas survival is expected does not change dramatically leading to a `universal' survival criterion $t_{\rm cool,mix}/\tilde t_{\rm cc}\lesssim 1$. 

\subsection{Dense mass growth rate versus increasing surface area}
\label{subsec:growth_rate_and_area}
In this section, we examine the dense mass growth rate \rits{in our simulations where cold clouds move through a hot wind subject to} 
continuous turbulent forcing 
and compare \rits{it to} the 
mass growth rate in laminar wind setups.

\begin{figure}
    \centering
    \includegraphics[width=\columnwidth]{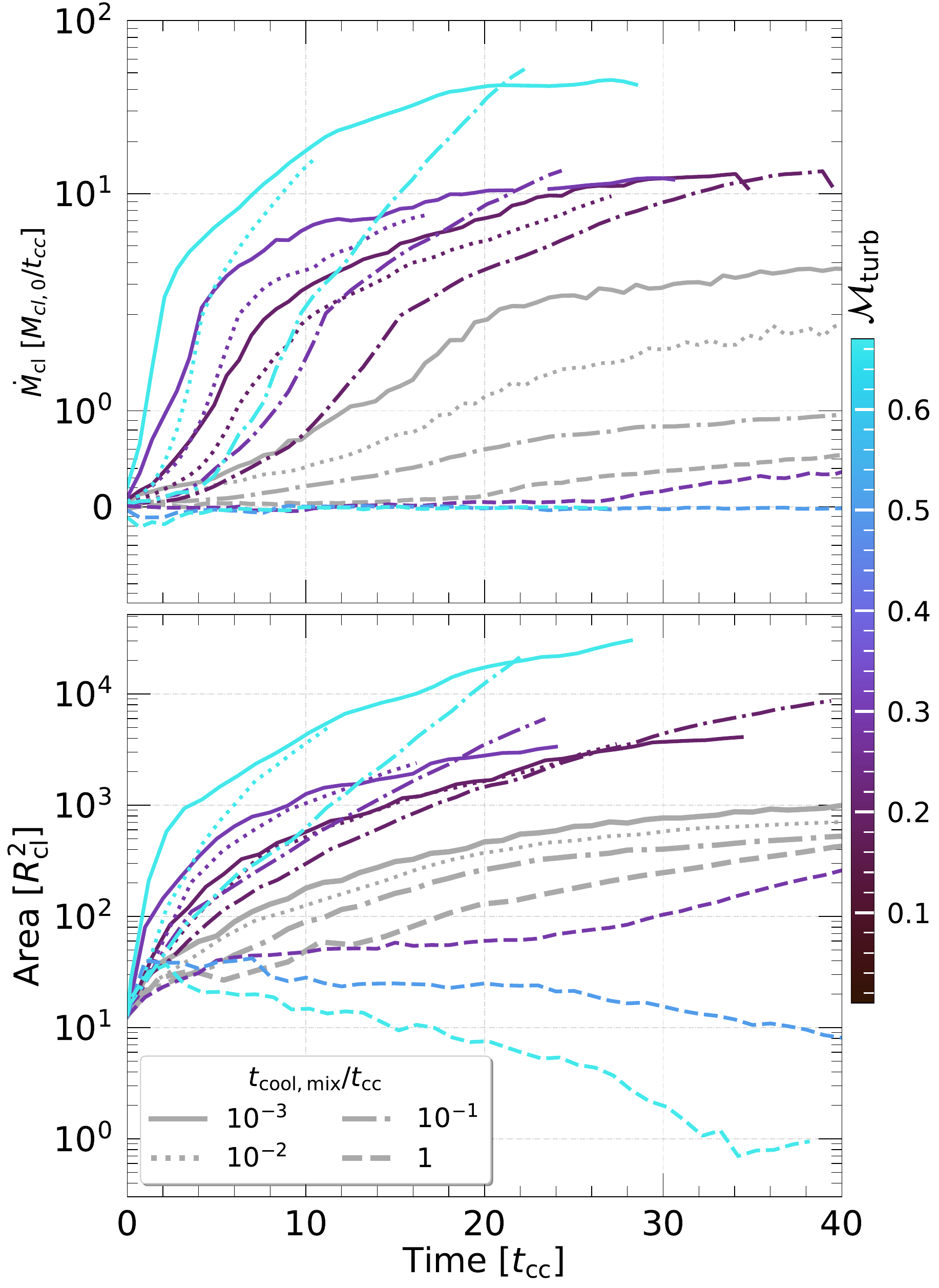}
    \caption{[\textit{Top panel}]: Mass growth rate for the fast and moderate cooling regimes with different strengths of turbulent forcing. Different linestyles are for different $t_{\rm cool,mix}/t_{\rm cc}$. 
    The line colors mark the evolution at different strengths of turbulent driving, as indicated in the colorbar. [\textit{Bottom panel}]: area as a function of time for an isosurface considered at temperature threshold $T_{\rm thres}=2\times T_{\rm cl}$). Continuous turbulent forcing enhances dense mass growth in the growth regime, by an order of magnitude. Similarly, the surface area available for mixing is enhanced by an order of magnitude for most of the growing cases.}  
    \label{fig:mass_area_growth}
\end{figure}

The top panel of Fig. \ref{fig:mass_area_growth} shows the mass growth rate $\dot{M}_{\rm cl}$ in units of $M_{\rm cl,0}/t_{\rm cc}$ ($M_{\rm cl,0}$ is the initial cloud mass and $t_{\rm cc}$ is the cloud crushing time). The line colors 
and line styles are similar to Fig. \ref{fig:mass_growth}.\footnote{\rits{To reduce noise in the estimation of the derivative of $M_{\rm cl}$, we apply boxcar smoothing with a kernel width of $0.01t_{\rm cc}$.}} 
The dense mass growth rate saturates at a higher value in comparison to a laminar wind with no turbulence (colored lines can be an order of magnitude above gray lines). 

A natural question that arises at this point is what determines the dense mass growth rate? As motivated in Section \ref{sec:intro}, the cold mass growth rate $\dot{M}_{\rm cl}$ can have two dominant contributions -- (i) an increasing effective area available for cooling (which we denote by $A_{\rm turb}$) and (ii) an increasing inflow velocity with which the mass accretes onto the cooling layer $v_{\rm mix}$. We try to disentangle the dominant mechanism among the two in the following few analyses.
We estimate $A_{\rm turb}$ in our simulation as the area of an isosurface at a temperature threshold \rits{of $T<2T_{\rm cl}$} (see \citealt{GronkeOh2018MNRAS.480L.111G} for a detailed discussion on similarity in the evolution of surface area at temperature thresholds like $2T_{\rm cl}$ and $T_{\rm mix}=\sqrt{T_{\rm cl}T_{\rm hot}}$ where cooling time is short).
In contrast, the mixing velocity $v_{\rm mix}$ can be derived numerically from the estimate of $A_{\rm turb}$ as $v_{\rm mix}=\dot{M}_{\rm cl}/\left(\rho_{\rm wind}A_{\rm turb}\right)$. 
The mixing velocity was found by high resolution turbulent mixing layer simulations \citet{TanOhGronke2021} as well a previous cloud-crushing work \citet{Gronke2020} to be of the order of the cold gas sound speed with an additional (in the `fast cooling' regime $-1/4$) dependence on the cooling time $v_{\rm mix}\sim c_{\rm s,cl}(t_{\rm cool,cl}/t_{\rm sc,cl})^{-1/4}$ \citep{Gronke2020,TanOhGronke2021,Fielding2020ApJ}.


\par The bottom panel of Fig. \ref{fig:mass_area_growth} shows the growth of the surface area of the mixing layer between the cloud and wind as a function of 
time. We find the surface area of an isosurface at temperature threshold $T_{\rm thres}=2T_{\rm cl}$ (where, $T_{\rm cl}$ is the initial temperature of the cloud and also the cooling floor) using the \texttt{marching cube} algorithm \citep{Lewiner2003}. 
In all the turbulent cases, we see an initial linear growth which is faster than the no turbulence runs shown in gray. This linear entrainment phase is followed by a slow increase in the area beyond $3-6\ t_{\rm cc}$. We find that the surface area available for 
cooling is enhanced by an order of magnitude in the fast cooling regime $t_{\rm cool,mix}/t_{\rm cc}$ with high turbulent forcing compared to a run with laminar flow (compare the cyan lines with the gray lines in the same linestyle).  

\begin{figure}
    \includegraphics[width=\columnwidth]{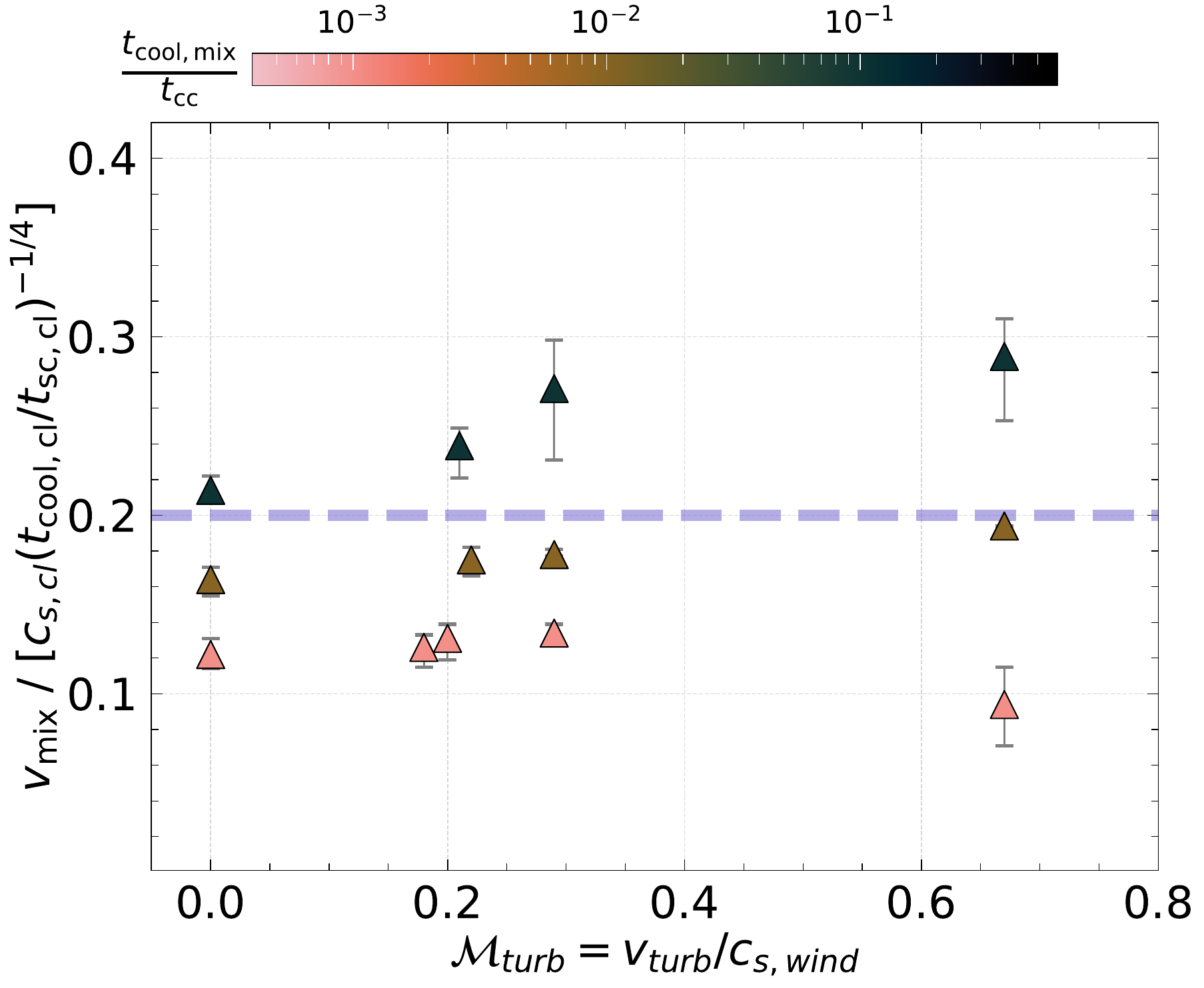}
    \caption{Derived mixing velocity $v_{\rm mix}$ (in units of $c_{\rm s, cl}/(t_{\rm cool,cl}/t_{\rm sc,cl})^{-1/4}$) as a function of turbulent Mach number $\mathcal{M}_{\rm turb}$ in various cooling regimes $t_{\rm cool, mix}/t_{\rm cc}$ (indicated in the colormap). The triangles indicate the average mixing velocity between $10-20\ t_{\rm cc}$, with error bars showing its range within the specified time. The derived mixing velocity is therefore close to $\sim 0.2$, as found in standard cloud crushing simulations. Driven turbulence does not significantly affect the mixing velocity with which dense mass condenses onto the surface area available for cooling.}
    \label{fig:mixed_velocity}
\end{figure}

Fig. \ref{fig:mixed_velocity} shows the evolution of the derived mixing velocity $v_{\rm mix}$ normalized by $c_{\rm s, cl}\left(t_{\rm cool,cl}/t_{\rm cc}\right)^{-1/4}$ ($c_{\rm s, cl}$ is the sound speed and $t_{\rm cool,cl}$ is the cooling time in the cold medium). Note, however, that while the exact value of $v_{\rm mix}$ is dependent on the estimate of $A_{\rm turb}$, which is not converged in the simulations, \maxg{the physically relevant surface area is an `effective' enclosing one which omits the small-scale wrinkles and (due to our resolution) is close to our estimate of $A_{\rm turb}$, which gives an upper limit on $v_{\rm mix}$ \citep[see extensive discussion about this in][]{Gronke2020}.}


\subsection{Faster entrainment}
\label{subsec:entrainment}
We now examine the entrainment of cold gas in galactic winds. While one might naively expect that increased turbulence would hinder cloud entrainment by disrupting the inflow of gas that is siphonig 
in from the hot \rits{gas onto the} cold interface, our results reveal the opposite trend: higher levels of wind turbulence actually facilitate the entrainment of cold gas.
This behavior is evident in Fig. \ref{fig:mass_growth}, where the curves corresponding to increasingly turbulent flows (colored lines with $\mathcal{M}_{\rm turb}$ according to the colormap) consistently lie below the laminar wind cases (gray lines), across a range of cooling regimes characterized by different values of the $t_{\rm cool, mix}/t_{\rm cc}$ ratio (indicated by different linestyles).

\begin{figure}
    \includegraphics[width=\columnwidth]{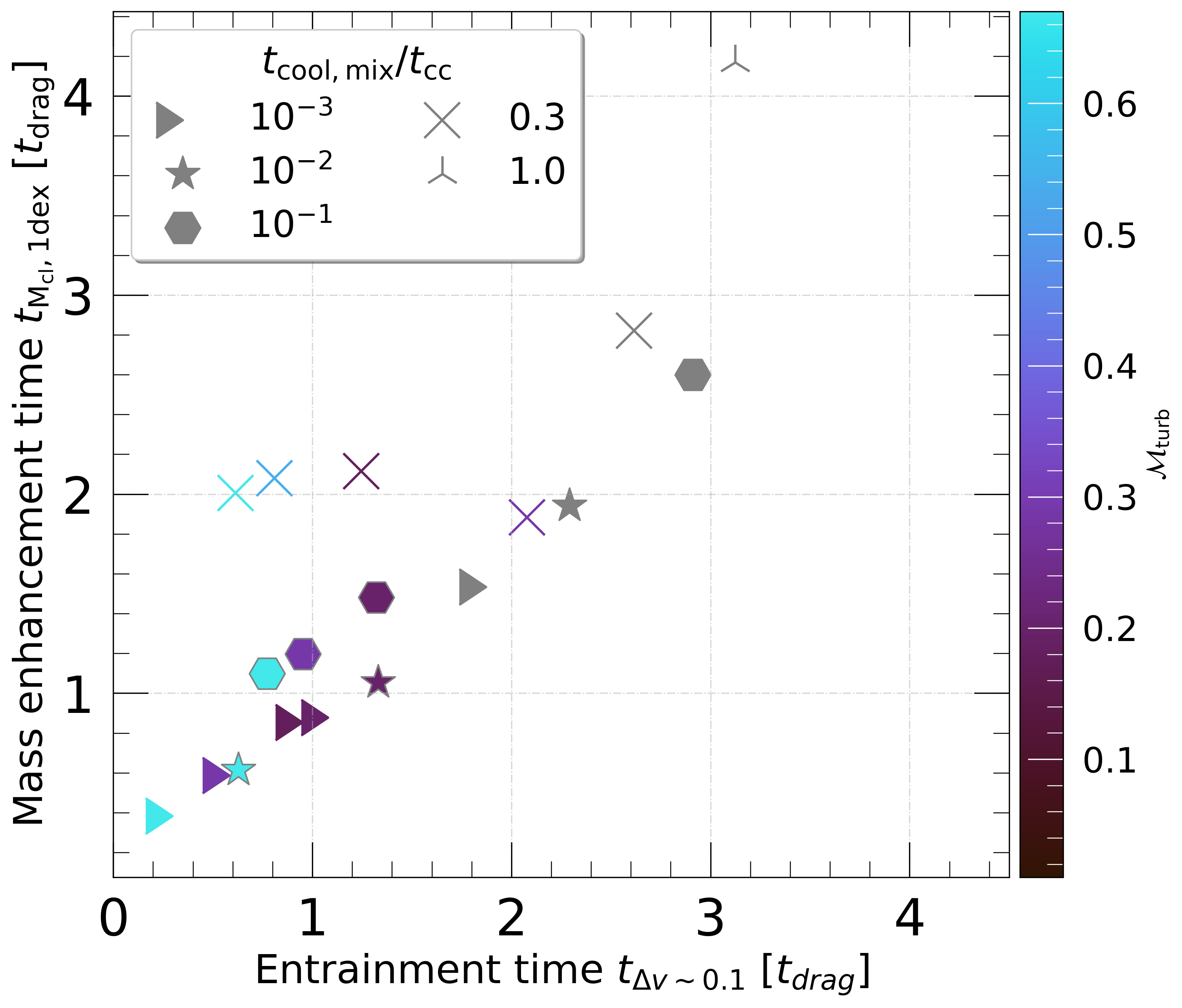}
    \caption{Correlation between mass enhancement time versus the entrainment time. A highly turbulent wind (higher $\mathcal{M}_{\rm turb}$) can entrain the cold clouds in a short time, while the mass is enhanced by an ever increasing area (see Fig. \ref{fig:mass_area_growth}).}
    \label{fig:correlation_massgrowth_entrainment}
\end{figure}

In Fig. \ref{fig:correlation_massgrowth_entrainment} we present the correlation between two key timescales: the mass enhancement time $t_{\rm M_{cl},1dex}$ and the cloud entrainment time $t_{\rm \Delta v \sim 0.1}$. 
We define the mass enhancement time\footnote{Note that this mass enhancement time is longer than the mass doubling time, which can be less than the drag time $t_{\rm drag}\sim \chi R_{\rm cl}/v_{\rm wind}$.} as the time required for cold mass to increase by a factor of 10 of its initial mass, while the entrainment time represents the time required \rits{for the relative velocity between the center of mass of the cloud and wind to reduce to} $10\%$ of the initial relative velocity i.e., $0.1v_{\rm wind,0}$. 
Both timescales in Fig. \ref{fig:correlation_massgrowth_entrainment} are normalized by the 
drag time $t_{\rm drag}\sim \chi R_{\rm cl}/v_{\rm wind}$ in cloud crushing simulations. Various markers are used for different $t_{\rm cool,mix}/t_{\rm cc}$, and they are colored according to the turbulent Mach number in the wind $\mathcal{M}_{\rm  turb}$, as indicated in the colormap. 
A strong, \rits{positive}, nearly monotonic correlation is evident between the two timescales in Fig. \ref{fig:correlation_massgrowth_entrainment}.

The cooling regime \rits{to which a cloud belongs (depending on its size)} influences the efficiency of cold gas entrainment. When cooling is slow,
($t_{\rm cool,mix}/t_{\rm cc}\sim 0.3, 1.0$ marked with crosses and tri-ups), although the cold mass could be enhanced by a factor of 10 on a few drag timescale, entrainment remains difficult.
In contrast, clouds in the fast cooling regime ($t_{\rm cool,mix}/t_{\rm cc}\sim 10^{-2}, 10^{-3}$ marked by triangles and stars) exhibit a markedly different behavior.
As turbulence in the wind increases, both the cold gas mass and the degree of entrainment improve significantly (see the magenta and cyan markers, with $\mathcal{M}_{\rm turb}$ indicated in the colormap), and this happens on timescales much shorter than the standard drag time.

The correlation between $t_{\rm M_{cl},1dex}$ and $t_{\rm \Delta v \sim 0.1}$ (especially in the fast cooling regime) is related to the enhanced mass growth rate which is a consequence of increasing area available for mixing of the hot and cold phases (Section \ref{subsec:growth_rate_and_area}).
Consequently, clouds experiencing rapid cooling and strong turbulence acquire momentum more effectively through increased mass entrainment. 


\section{Discussion}\label{sec:discussion}
\subsection{Growth of turbulent tails and their morphology}\label{subsec:morphology_with_turb}

\begin{figure}
    \centering
    \includegraphics[width=\columnwidth]{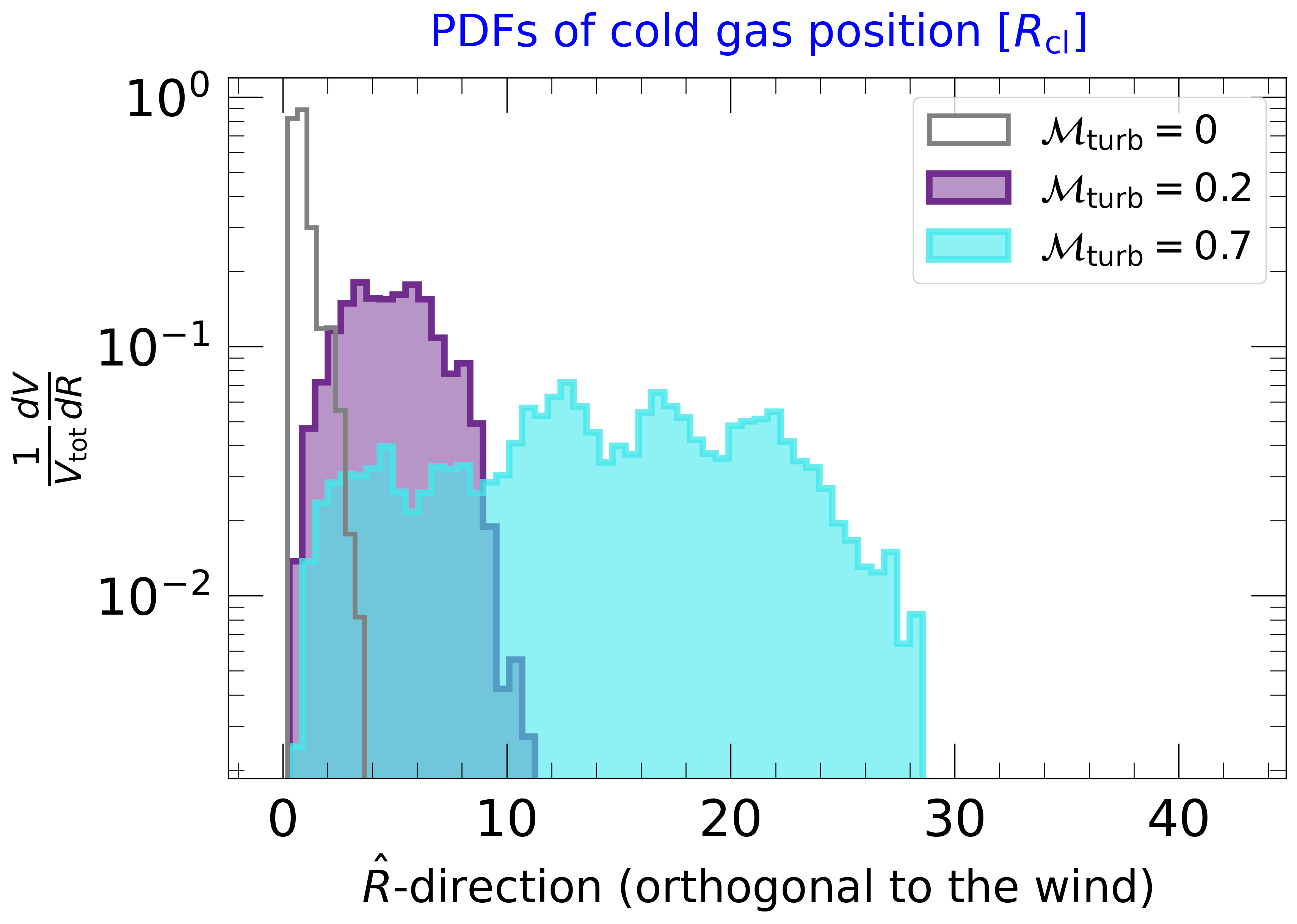}
    \includegraphics[width=\columnwidth]{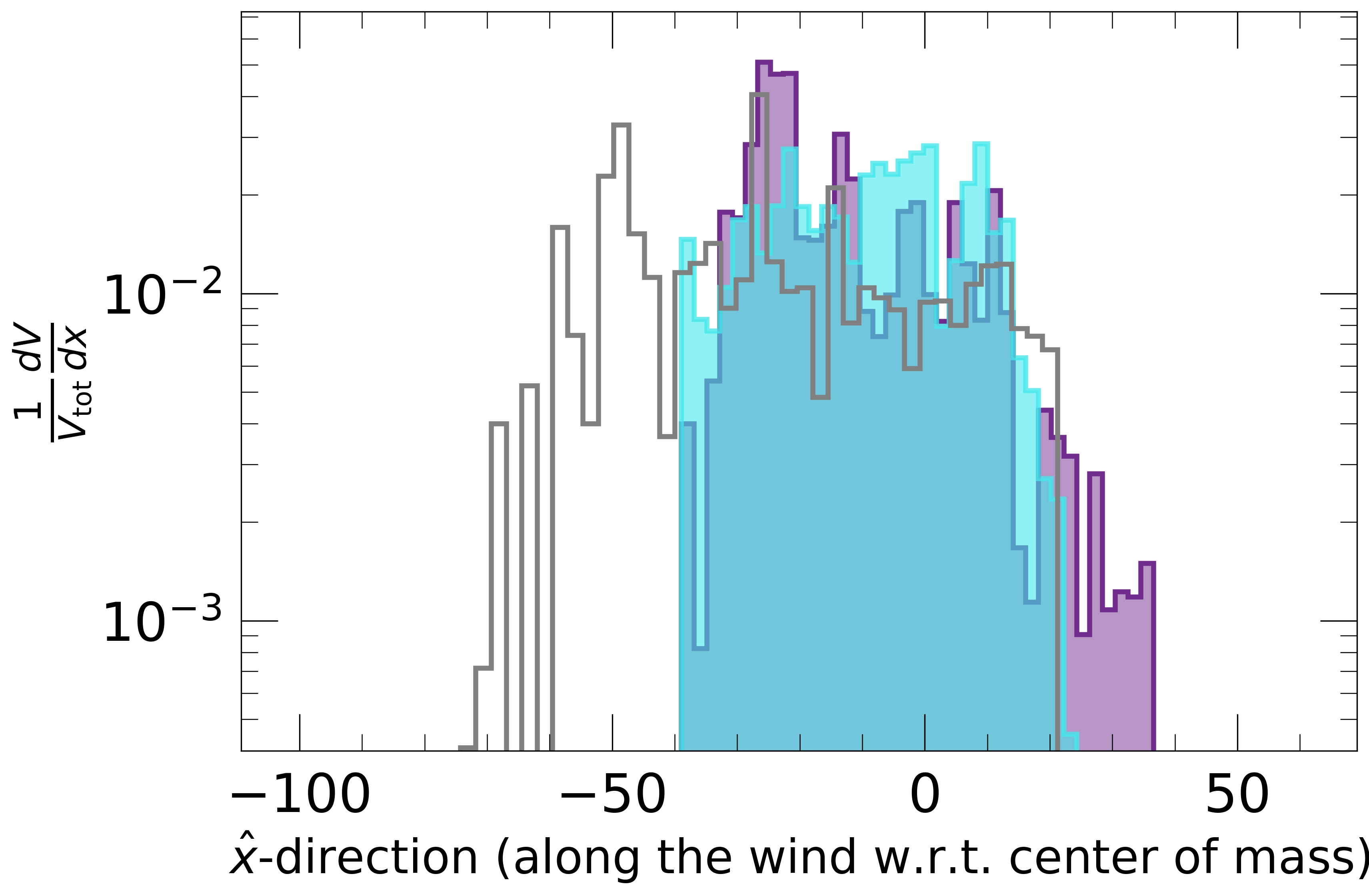}
    \caption{
    \rits{The spatial distribution of cold gas in turbulent and laminar winds. Both panels show histograms of dense cell positions at $10\ t_{\rm cc}$ for a simulation in the growth regime ($t_{\rm cool, mix}/t_{\rm cc}=0.1$).}
    [\textit{Top panel:}] 
    \rits{Distribution} in the orthogonal direction \rits{($\hat{R}$; representing the radial distance from the initial cloud center)} for various turbulent Mach numbers $\mathcal{M}_{\rm turb}$. 
    [\textit{Bottom panel:}] The \rits{distribution} 
    in the direction of relative motion \rits{($\hat{x}$-direction) measured} with respect to the center of mass of the cloud.
    In comparison to a laminar wind with no turbulence (\rits{\textit{gray}} histogram) which is elongated along the direction of relative motion with narrow width in orthogonal direction, we find that \rits{increasing turbulence (\textit{magenta} and \textit{cyan} histograms)} makes the cold mass grow into 
    \rits{an isotropic cloud complex} with a much shorter longitudinal extent (also see Fig. \ref{fig:overview_cc_turb}).
    }
    \label{fig:dense_histogram}
\end{figure}

Fig. \ref{fig:dense_histogram} shows the volume-weighted, normalized histograms of the distribution of dense gas in the direction parallel (bottom panel) and perpendicular (top panel) to the wind, for various turbulent Mach numbers $\mathcal{M}_{\rm turb}$ in the hot phase at $10\ t_{\rm cc}$. The analysis is restricted to the growth regime, with a ratio of the cooling time of mixed gas and the cloud-crushing time $t_{\rm cool, mix}/t_{\rm cc}=0.1$. Only computational cells with densities exceeding one-third of the initial cloud density 
($\rho>\rho_{\rm cl,0}/3$) are included in the histograms to isolate the cold, dense phase. We then define the orthogonal spread $R=\sqrt{y^2+z^2}$ within a cylinder that is centered at the cloud's initial position and whose axis is aligned with the direction of relative motion.

Compared to simulations with a laminar wind, the presence of external turbulence significantly alters the morphology of the cold gas. In turbulent environments, the dense material evolves into a more extended structure, with the transverse (orthogonal) spread of the cold gas reaching up to $20 R_{\rm cl}$ for $\mathcal{M}_{\rm turb}\sim 0.7$ in the top panel of Fig. \ref{fig:dense_histogram} (cf. Fig. \ref{fig:overview_cc_turb}). 

The bottom panel of Fig. \ref{fig:dense_histogram} shows the distribution of dense gas positions along the direction of the wind $\hat{x}$, with respect to the center of mass of the cloud at $10\ t_{\rm cc}$. With increasing turbulence, the elongated comet-like tail observed in the laminar case (the gray histogram for $\mathcal{M}_{\rm turb}=0$) becomes less pronounced (the cyan histogram with $\mathcal{M}_{\rm turb}=0.7$). Simultaneously, the cloud exhibits increased lateral dispersion in the top panel, leading to a larger cross-sectional area. \rits{All these} suggest that external turbulence not only enhances cold mass growth but also modifies the geometry of the cloud, promoting a more isotropic distribution of the dense phase.
 

\begin{figure}
    \centering
    \includegraphics[width=\columnwidth]{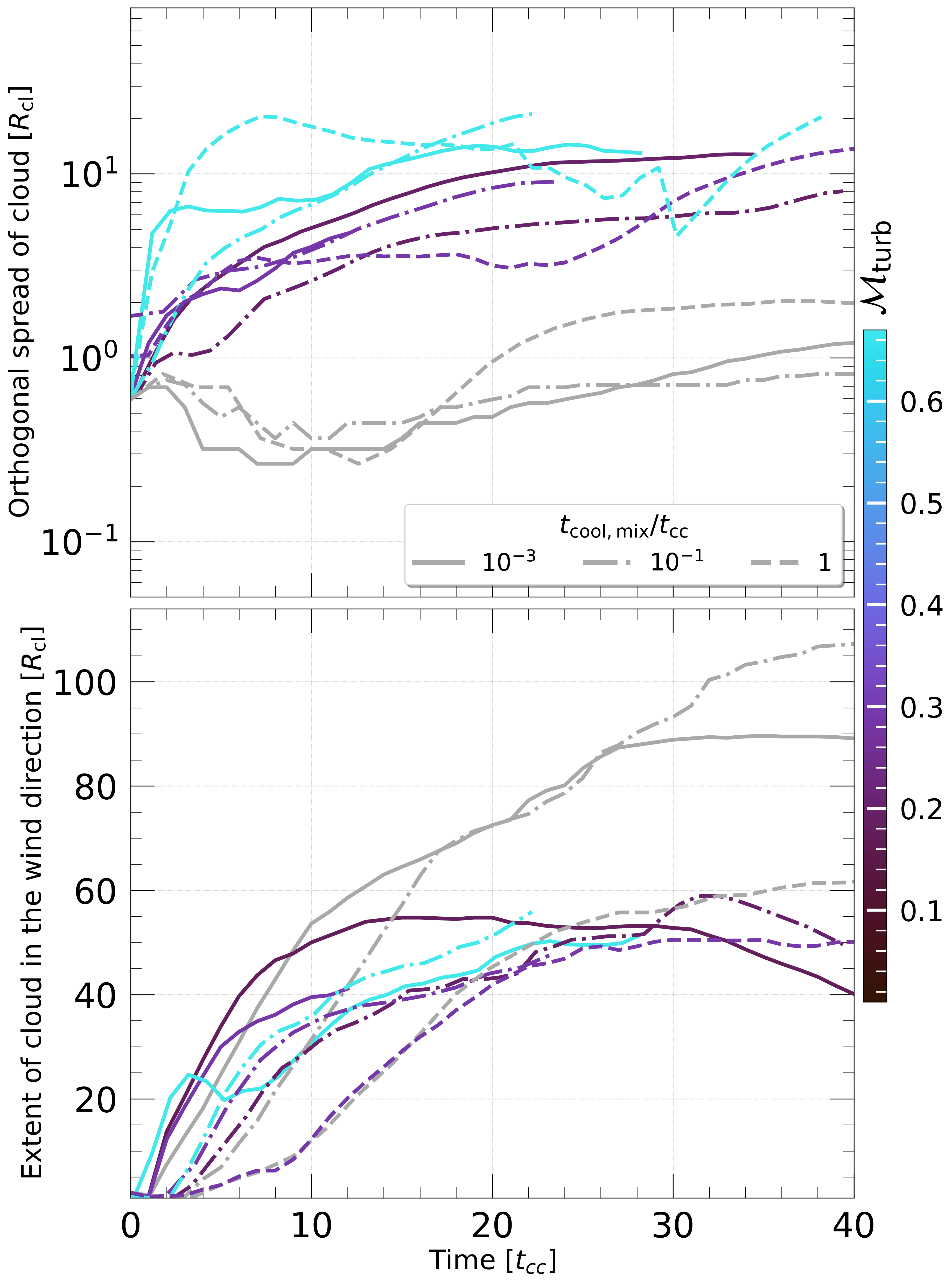}
    \caption{
    \rits{Evolution of cold gas spread and tail length in the presence of continuous turbulent forcing in cloud crushing simulations.
    [\textit{Top panel}]: Orthogonal spread of cold gas (in units of the initial cloud radius $R_{\rm cl}$) perpendicular to the wind direction ($\hat{R}$) 
    for simulations with varying turbulent forcing, as indicated by the Mach number $\mathcal{M}_{\rm turb}$ in the colormap. 
    The extent of cloud is determined from the lateral spread of dense gas as described in \S~\ref{subsec:morphology_with_turb}. 
    [\textit{Bottom panel}]: Longitudinal spread of cold gas (i.e., the tail length) along the wind direction ($\hat{x}$). 
    The figure highlights how increasing turbulence produces shorter, but wider, cold gas tails compared to standard laminar wind, which typically exhibit long, filamentary tails.}
    }
    \label{fig:tail_length}
\end{figure}

Fig. \ref{fig:tail_length} shows the evolution of the extent of cold gas in the orthogonal direction in the top panel and in the direction of the wind $\hat{x}$ in the lower panel. The extent of the cloud is expressed in terms of the initial cloud size $R_{\rm cl}$. Cells with a density greater than one-third of the initial cloud density ($\rho>\rho_{\rm cl,0}/3$) are considered in the analysis and the longitudinal extent is measured as the difference between the 90 and 10-percentiles of the position of dense gas along the direction of the wind $\hat{x}$. Similarly, the orthogonal extent is determined from the difference between 90 and 10 percentiles of the spread of the cloud along $R$, the radial extent of the cloud (computed in a cylinder centered at the initial cloud position and axis aligned with the direction of relative motion). 
Compared to the laminar case (gray lines), higher turbulence significantly stretches the cloud orthogonally (colored lines in the top panel) while reducing the longitudinal extent by a factor of a few (colored lines in the bottom panel). Overall, we find that, compared to the long cometary tails formed in laminar winds, turbulence increases the cloud's orthogonal extent, resulting in a significantly more isotropic and compact cloud complex. 
This morphological change can have important observational implications, for instance, cold gas would have a substantially larger area covering fraction in turbulent galactic winds.

\subsection{Turbulent \textit{enhancement} of cold gas}
\label{subsec:turbulent_enhancement_of_cold_gas}
While vanilla `cloud crushing' employ a laminar wind \citep[e.g.][]{Klein1994ApJ}, realistic galactic winds are highly turbulent -- as shown through observations (eg, M82 in H$\alpha$: photometry \citealt{Shopbell1998ApJ}, HST imaging \citealt{Mutchler_2007}; X-ray: Chandra observations \cite{Lopez_2020}; PAH: with JWST \citealt{Fisher2025}) and simulations \citep{Creasey2013, Schneider_2018, Schneider_2020, Kim2018ApJ, Vijayan_2020, Tan2023}. 
This discrepancy raises doubts for the applicability of `wind tunnel' simulations, in particular, if the commonly studied `survival criterion' \citep{GronkeOh2018MNRAS.480L.111G, Kanjilal2021, Abruzzo2022ApJ} holds as additional turbulence might lead to a rapid mixing of cold gas, modifying the survival criterion.

In spite of these expectations, our numerical experiments show that driven turbulence does not change the survival dramatically, and we suggest a survival criterion incorporating the effect of additional turbulence (cf. \S~\ref{subsec:growth_cc}, Eq.~\ref{eq:tcc_tilde}). In fact, instead of facilitated destruction we find that turbulence in the wind enhances the growth of cold gas.
While somewhat surprising, this enhancement of cold gas production is easy to understand. 
As ``mixing is stretching enhanced diffusion'' \citep{VillermauxAnnurev2019} the additional turbulence can lead to either more `stretching', i.e., an accelerated growth in cold gas surface area, or to more (turbulent) diffusion -- both of which would ultimately lead to more mixed gas and thus an increase in $\dot M_{\rm cl}$. In \S~\ref{subsec:entrainment}, we could show that for cold clouds entrained in a turbulent wind, the former effect is dominant (we will discuss this further below).


\begin{figure}
    \includegraphics[width=\columnwidth]{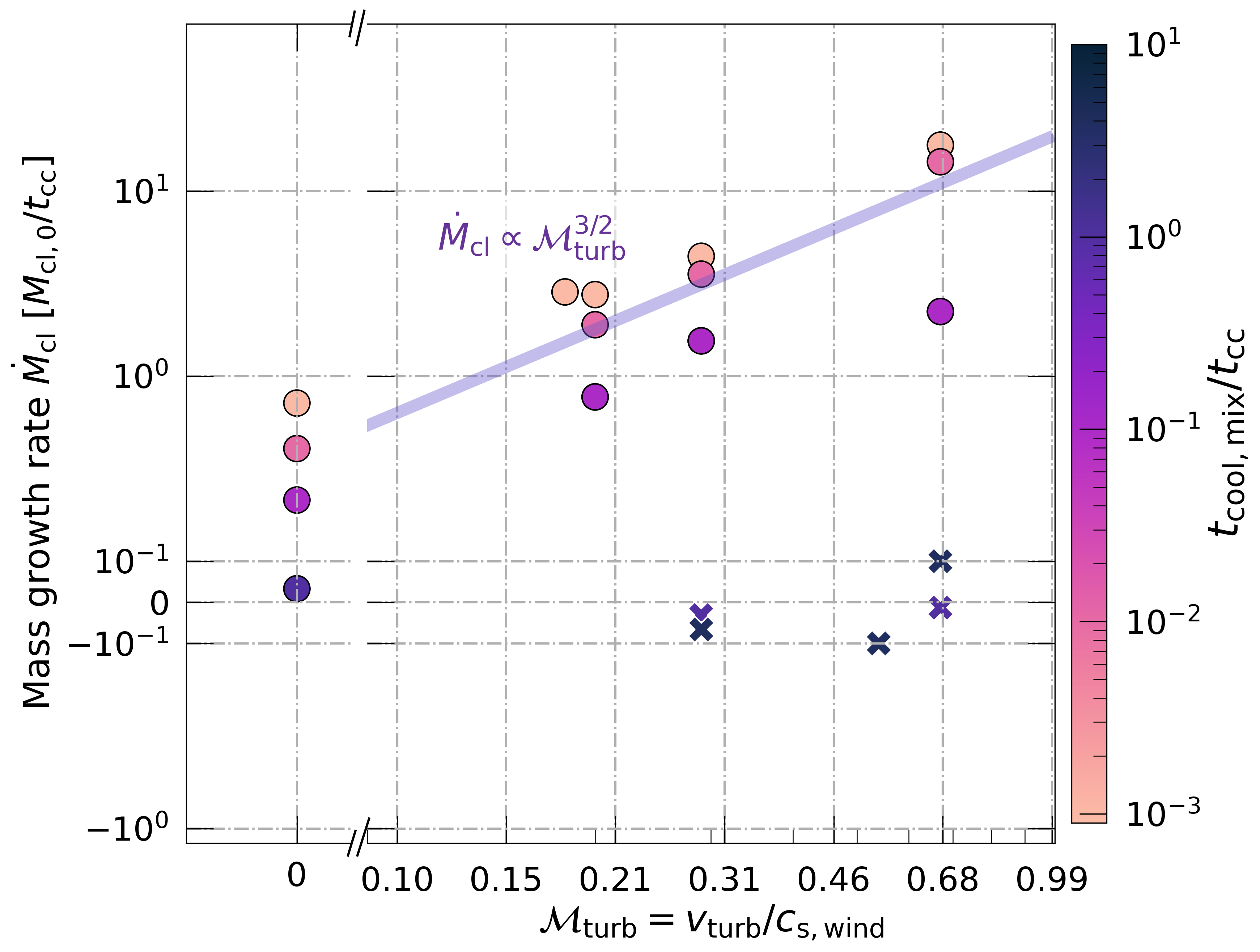}
        \caption{Overview plot for the cold mass growth rate $\dot{M_{\rm cl}}$ (in units of $M_{\rm cl,0}/t_{\rm cc}$; and measured at $10t_{\rm cc}$) versus turbulent Mach number $\mathcal{M}_{\rm turb}$. The circles denote cloud growth while the cross marks denote cloud destruction (we denote a cloud to be destroyed if dense mass falls below 10\% of initial value). The points are color-coded according to the value of $t_{\rm cool,mix}/t_{\rm cc}$ for the simulations, as indicated in the colorbar.
        Across simulations with varying radiative cooling strengths, we find a clear correlation between cold gas growth rate and the turbulent Mach number in the hot phase approximately scaling as $\dot{M}_{\rm cl}\propto \mathcal{M}_{\rm turb}^{3/2}$.
        }
    \label{fig:overview_growth_rate}
\end{figure}

In Fig. \ref{fig:overview_growth_rate}, we plot the evolution of cold gas growth rate (an average value beyond $10\ t_{\rm cc}$ in the top panel of Fig. \ref{fig:mass_area_growth}) as a function of turbulent Mach number in the hot phase. The crosses represent the destroyed clouds in contrast to the circled points indicating the growing ones.
Depending on the strength of radiative cooling (as indicated in the colorbar), we find, in the fast cooling regime, a clear correlation between
cold mass growth rate and the turbulent Mach number.
We can identify the growth of cold mass $\dot{M}_{\rm cl}$ as proportional to $\mathcal{M}_{\rm turb}^{3/2}$. 
\rits{This trend highlights the role of turbulence in enhancing the mixing and subsequent condensation of cold gas in cloud–wind interactions.}

\begin{figure}
    \includegraphics[width=\columnwidth]{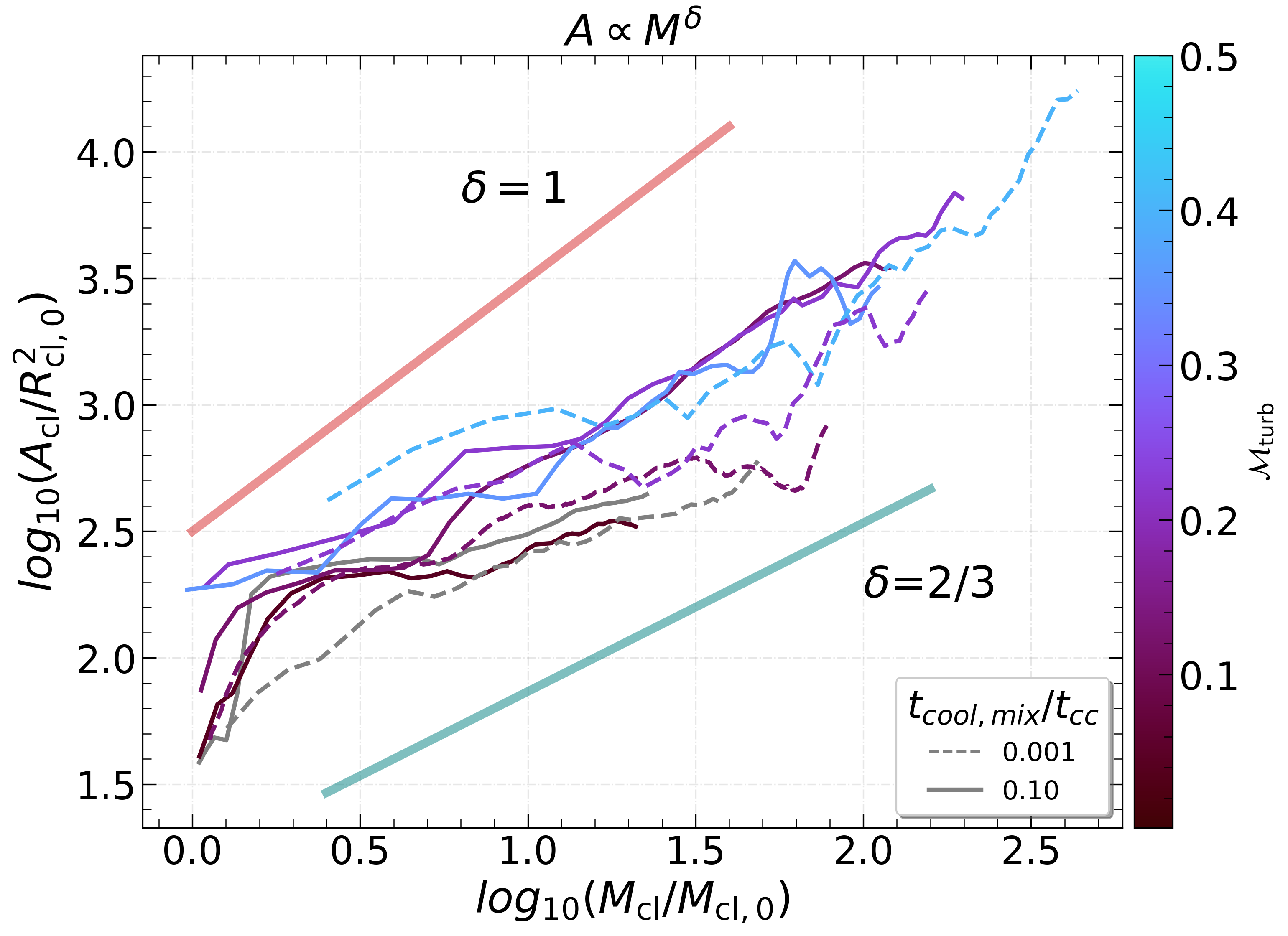}
    \caption{
    \rits{Relation between cloud area $A_{\rm cl} $(in units of $R_{\rm cl}^2$) and cold gas mass $M_{\rm cl}$ (in units of initial cloud mass $M_{\rm cl,0}$) for simulations with varying turbulent forcing, as indicated by the turbulent Mach number in the colorbar. Thick cyan and orange lines show reference scalings $A_{\rm cl}\propto M_{\rm cl}^\delta$. For most runs with driven turbulence, the evolution follows a slope of $\delta \approx2/3$, indicating that the area available for cooling of mixed gas roughly tracks the expected geometric expression.
    } 
    }
    \label{fig:mass_area_relation}
\end{figure}

For the cases that show an enhanced cold mass, we identified in Section \ref{subsec:growth_rate_and_area} that the inflow velocity is unchanged, but instead the larger $\dot M_{\rm cl}$ is due to the larger surface area through which the cold gas can accrete mass.
Fig.~\ref{fig:mass_area_relation} shows the evolution of the area in units of $R_{\rm cl,0}^2$ with cold gas mass $M_{\rm cl}$ in units of initial cloud mass $M_{\rm cl,0}$. Thick lines in cyan and orange, mark the relation $A_{\rm cl}\propto M_{\rm cl}^\delta$. For a long time, most of the runs with driven turbulence (as indicated in the strength of turbulent Mach number in the color bar), follow a slope of 2/3, suggesting that the area available for cooling of mixed gas produced by turbulent driving remains close to the expected geometric expansion. For the cyan curve in dashed linestyle, we do see a transition to a slope close to 1, which can be due to the fractal nature of the cooling surface in extremely strong cooling.

This increased mass growth rate has several implications. Foremost, it affects, of course, the morphology of the cold gas (as discussed in Section ~\ref{subsec:morphology_with_turb}) with clouds being `puffier' and showing less cometary and, elongated tails. Furthermore, the linked larger momentum transfer from the hot to the cold medium implies faster entrainment of the clouds as shown in \S~\ref{subsec:entrainment}. In fact, we find entrainment times to drop from 2-3 $t_{\rm drag}$ (in the vanilla cloud-crushing) to $\sim0.2\ t_{\rm drag}$ with (high) turbulence. This would correspond to a cloud travel distance of $20\ R_{\rm cl}$, i.e., $\sim 70$\,pc (using the fiducial parameters of Eq.~\ref{eq:survival_threshold_size}) in order to be fully entrained. Both, the cold gas morphology as well as the velocity gradient of cold gas can potentially be observed in nearby galaxies (which we discuss further in \S~\ref{subsec:observational_implications}).

\subsection{Implications for larger scale simulations} 

Our findings have several implications for larger scale, cosmological simulations and semi-analytical/empirical models of galactic winds. 

First, akin to \citet{GronkeOh2018MNRAS.480L.111G}, we can write the survival criterion Eq. \ref{eq:tcc_tilde} as a geometrical criterion as
\begin{equation}
    R_{\rm cl}\gtrsim 3.4\ \text{pc}\
    \left(\frac{\chi}{100}\right)
    \frac{T_{\rm cl,4}^{5/2}\mathcal{M}_{\rm wind}}{(P_3/k_{\rm B})\Lambda_{\rm mix, -21.3}}\
    \sqrt{1+\left(\frac{\mathcal{M}_{\rm turb, 0.3}}{\mathcal{M}_{\rm wind}}\right)^2}\ ,
    \label{eq:survival_threshold_size}
\end{equation}
where $T_{\rm cl,4}\simeq T_{\rm }/(10^4\rm K)$, $P_3/k_{\rm B}\simeq nT / (10^3\ \rm cm^{-3} \ K)$,
$\Lambda_{\rm mix, -21.3}\simeq \Lambda ( T_{\rm mix} ) / (10^{-21.3}\rm \ erg \ cm^{-3}\ s^{-1})$, $\mathcal{M}_{\rm wind}$ is the Mach number of the hot wind, $\mathcal{M}_{\rm turb,0.3}\simeq \mathcal{M}_{\rm turb}/{0.3}$ is the turbulent Mach number in the hot wind and $\chi$ is the density contrast of the cloud with the wind.
This scale -- above which we expect cold clouds in a turbulent wind to survive, is orders of magnitude larger than the (halo) resolution of many cosmological simulations (likely leading to non-convergence in their cold gas properties, cf. discussion in \citealp{vandeVoort2021, Hummels_2019, Peeples_2019}).
This means that \rits{the dynamics of} many of the clouds we expect from this study to survive \rits{would not be accounted for}, and thus transport mass and metals into the CGM (or back onto the disk; cf. \citealp{Fox_2019, PerouxHowk2020}) would not exist in such simulations.

In this work, we also found that the growth rate of entrained clouds critically depends on the turbulent properties of the wind with $\sim 1$ dex larger mass transfer rates found for a turbulent versus a purely laminar wind (cf. \S~\ref{subsec:growth_cc}). Thus, in order to obtain, e.g., the right amount of cold gas mass loading rates and the correct exchange rates between galactic disks and their associated halos, this needs to be taken into account and modeled correctly. Obtaining the correct wind turbulence is non-trivial as in resolved simulations it requires to be able to resolve not only the stirring length-scale but also to follow the turbulent cascade down to much smaller scales of molecular diffusion to produce an inertial range. 
This \rits{can be} 
particularly challenging for adaptive techniques employed \rits{by conventional cosmological simulations,}
which usually have orders of magnitude worse spatial resolution in the hot medium (see, e.g., discussion in \citealp{Vazza2010} related to turbulence in the ICM). For galactic winds, e.g., \citealt{Smith2018MNRAS} 
achieved a mass resolution of 20$M_{\odot}$ which corresponds to $\sim 10$ pc and $\sim 50$ pc in the cold and hot medium, respectively (similar numbers were achieved by other studies; e.g., \citealt{Steinwandel_2024}). 
This is not only significantly larger than required to resolve the critical `survival length scale' (Eq.~\ref{eq:survival_threshold_size}) but it is yet unclear how this affects the turbulence in the hot medium, and the associated mass transfer rate.

Furthermore, the reason galactic winds are turbulent is intricately linked to \rits{the details at the site of} their launching, 
for instance, their launching energy and location, 
the initial `resistance' due to close cold gas structure in the ISM.
Our results show that the multiphase structure and loading factors of galactic winds are strongly dependent on their turbulent properties, and accurate modeling of these winds requires a precise representation of the launching conditions and the detailed structure of the ISM.

\rits{Energy transport and the thermodynamics of the hot and cold gas interaction is dominated by the highly efficient radiative cooling of warm turbulent gas in the mixing layers. Small scale idealized simulations have routinely found that emission is dominated by the parsec to sub-parsec scale mixing layers \citep{LiHopkins2020, Gronke2020,TanOhGronke2021, Kanjilal2021} which are either poorly resolved or completely unresolved in cosmological simulations. Further work is needed to accurately model the mass and energy transport, and radiative loses in these mixing layers, validate them with small-scale simulations, and bridge the existing gap in cosmological simulations between unresolved small scale physics and their 
effect on large resolved scales. For instance, the continuous mass growth across the turbulent boundary layers formed around entrained cold clouds in our simulations can be either due pressure gradient subsonically siphoning in new, hot gas -- as suggested by \citealt{Dutta2022} (also see \citealp{Kanjilal2021}) or due to turbulent mixing \citep[see, e.g.,][]{Fielding2020ApJ,TanOhGronke2021,TanOh2021}.
Note that Appendix \ref{app:differential emission} and Fig. \ref{fig:differential_emission} we show tentative evidence that local subsonic cooling flows develop across the clouds in our simulation once they are sufficiently entrained. 
}

Apart from `resolved' hydrodynamical simulations, this dependence on the extrinsic turbulent properties of the mass transfer rates has also implications for sub-grid models \citep{PhEW_Huang2020, Smith2024arXiv,Hitesh2024arXiv241203751D} or analytical models of galactic winds \citep{FieldingBryan2022,Nguyen2023MNRAS.518L..87N,NikolisGronke2024}.
Thus far, only the turbulence due to the Kelvin-Helmholtz instability arising because of the relative motion between the hot and cold material has been considered, thus highly 
underestimating the actual mass transfer 
from the hot to the cold phase (as shown in \S~\ref{subsec:growth_rate_and_area}).





\subsection{Observational implications}\label{subsec:observational_implications}
\begin{figure*}
    \centering
    \includegraphics[width=\textwidth]{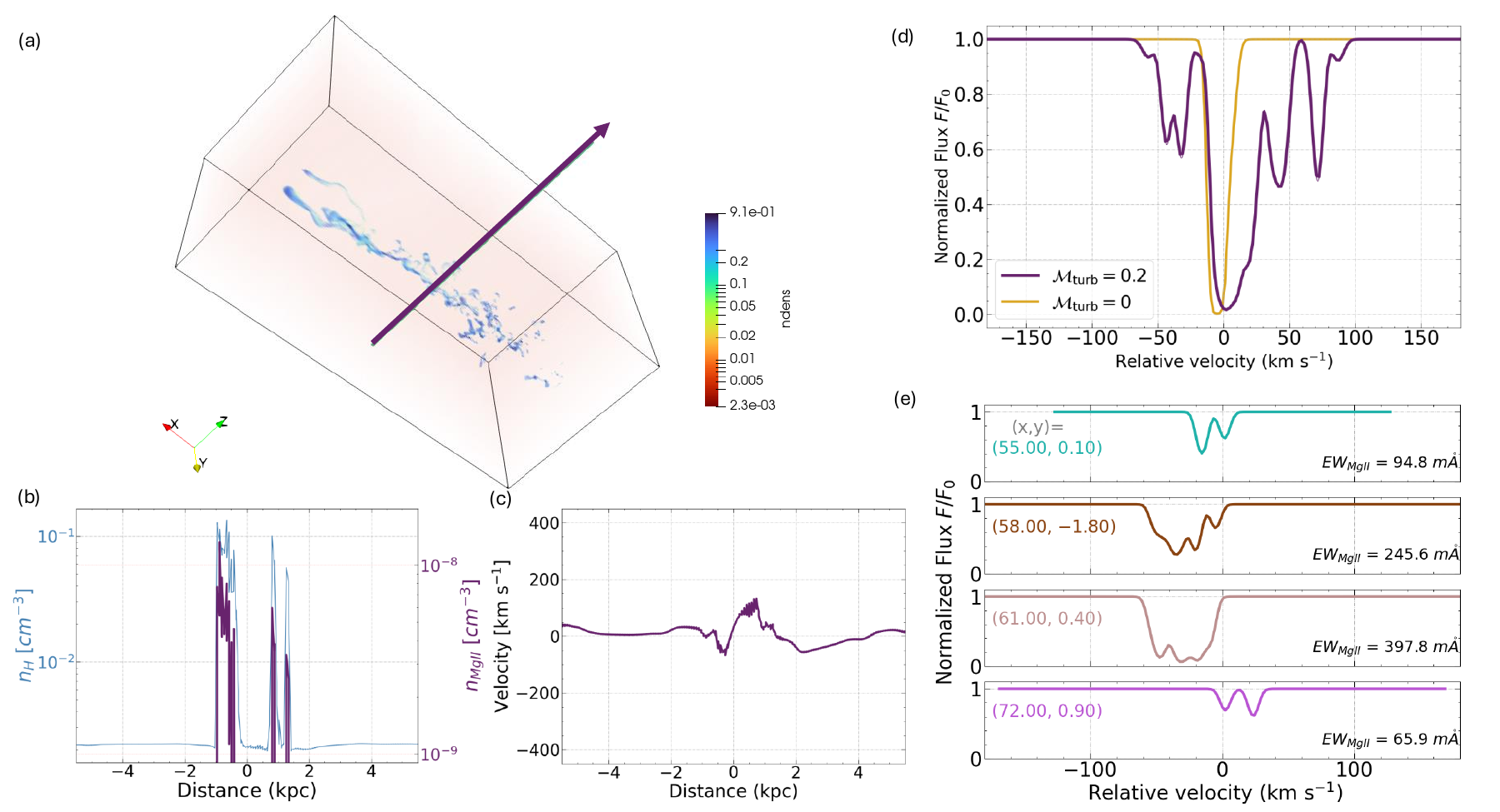}
    \caption{The column of MgII through a randomly chosen line of sight for $t_{\rm cool,mix}/t_{\rm cc}=10^{-1}$ and $\rm M_{\rm turb}=0.2$. 
    The top-left panel (a) shows the volume-rendered view with the chosen sightline intersecting three \rits{cloud complexes} 
    (visible as overdensities in panel b).
    Panel (b) shows the Hydrogen number density (in blue) along with that of MgII in magenta. The velocity field along the direction of the line of sight ($\hat{z}$) is shown in panel (c) . 
    The resultant absorption profile is shown on the top right corner in panel (d) in magenta color. With turbulence, the absorption spectrum is much broader ($\rm EW_{\lambda 2796}=550 \ m  \mathring{A}$) compared to the cloud in a uniform wind (shown in yellow line; $\rm EW_{\lambda 2796}=176\ m \mathring{A}$), and traces multiple kinematic components.
    In the bottom right panel (e), we show the absorption profile for MgII through multiple sightlines (passing through the annotated x,y coordinate along z-direction) in the cloud's tail. The equivalent width for MgII $2796 \ \rm \mathring{A}$ transition is indicated in each panel.
    }
    \label{fig:cloud_MgII_columns}
\end{figure*}

\begin{figure}
    \centering
    \includegraphics[width=0.96\columnwidth]{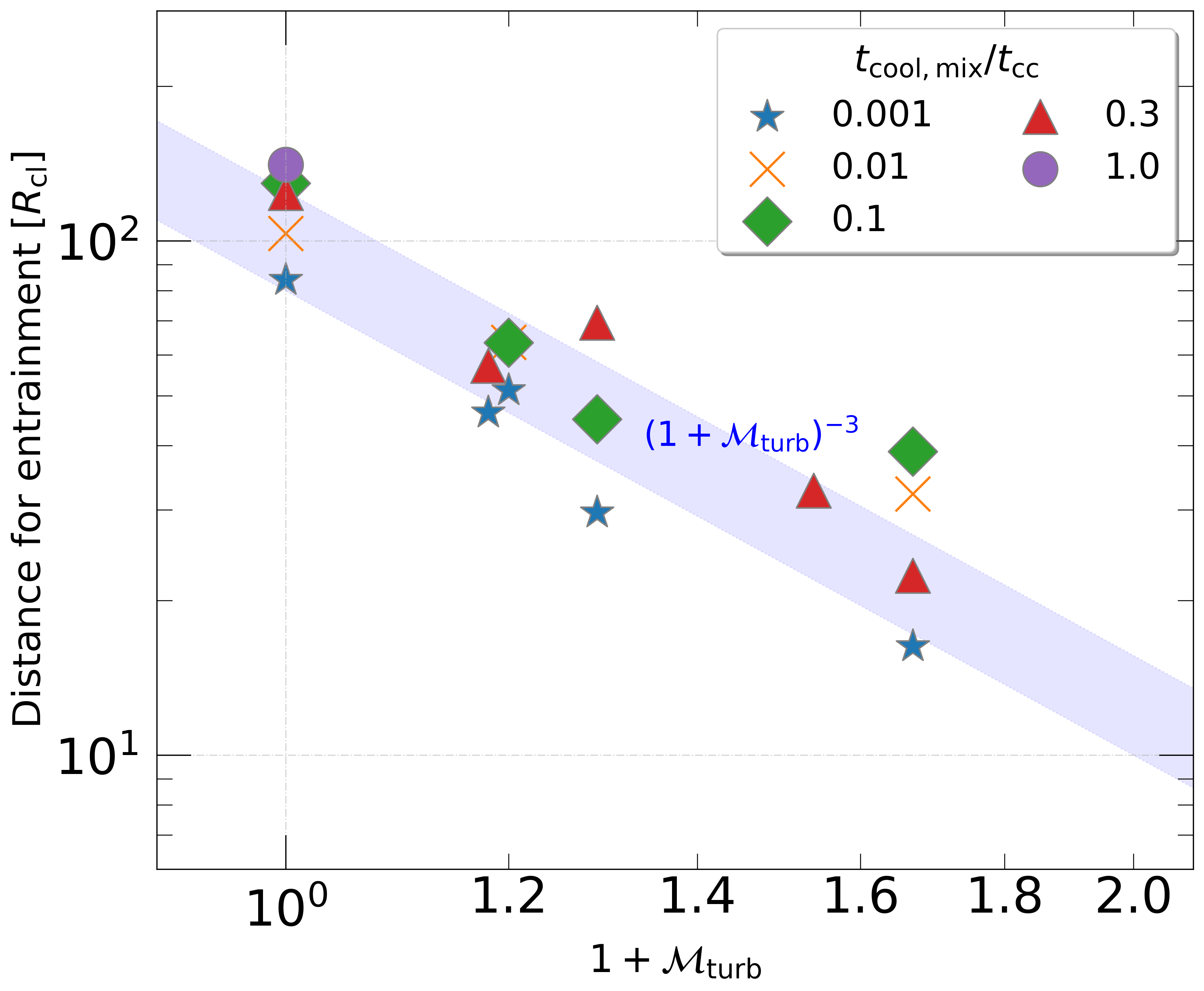}
    \caption{Effective entrainment distance defined as the distance the cloud travel when the relative velocity reaches $\Delta v / v_{\rm wind}\sim 0.1$ against turbulent Mach number. The different symbols indicate different cooling efficiencies and the blue band marks the scaling $\propto (1+\mathcal{M}_{\rm turb})^{-3}$ which the numerical results seem to follow.}
    \label{fig:Drag time}
\end{figure}
Global simulations, such as those of \citet{Schneider_2020, Creasey2013, Vijayan_2020}, highlight that galactic outflows can generate substantial turbulence down the wind. These simulations suggest that turbulence is a natural byproduct of outflows, that contributes to the mixing and interaction of different gas phases \citep[cf. fig. 21 of][]{Schneider_2020}.
Numerous examples in the literature indicate that the turbulent Mach number in the hot phase typically remains close to subsonic values (\citealt{Rudie_2019,ChenQu_2023}). This subsonic turbulence is crucial for understanding the dynamics of galactic winds as well as the intracluster medium (ICM) and the evolution of galaxies within these clusters.

Our findings suggest that turbulent mixing can boost the rate of cold mass growth by an order of magnitude (see Fig. \ref{fig:mass_area_growth}). This is particularly interesting, as this would imply that the radiative luminosity in the cold phase $\mathcal{L}$ can be boosted by $\sim 1$ dex (\citealt{Fabian1994ARA&A..32..277F}) 
This increase is particularly significant for interpreting observations of galactic winds and understanding the energy distribution within these systems. The morphology of these clouds combined with line-widths measurements, such as those seen in M82's filamentary PAH and CO emission (\citealt{Fisher2025}), 
can be used to constrain the level and nature of turbulence within the wind.


\rits{As an illustration of the observational implications of turbulence on cold gas growth, we compare the MgII columns and EW from our simulations with and without turbulence. }
The top left panel in Fig. \ref{fig:cloud_MgII_columns} shows a volume rendering of one of our simulations with $t_{\rm cool,mix}/t_{\rm cc}=10^{-1}$ and turbulent Mach number $\mathcal{M}_{\rm turb}=0.2$ in the hot phase at time $t\sim 10t_{\rm cc}$. We use \href{https://github.com/dutta-alankar/AstroPlasma}{\texttt{AstroPlasma}} \citep{Dutta2024} for calculating the MgII density for each cell in this snapshot. \texttt{AstroPlasma} uses pre-computed \texttt{Cloudy} \citep{Ferland2017} equilibrium \rits{photo+collisional} ionization models to estimate ion populations in the presence of a \cite{HaardtMadau2012ApJ} \rits{extragalactic UV background at redshift 0}. We choose a random line of sight passing through the tail (along the $\hat{z}$-direction, \rits{which is orthogonal to the wind}) to estimate the absorption flux at MgII 2796 $\mathring{A}$, typically seen in quasar spectra (\citealt{Draine2011piim.book}).
The resultant absorption profile is shown in the purple curve on the top right corner \rits{of Fig. \ref{fig:cloud_MgII_columns} (panel d),} and is much broader compared to the cloud in a uniform wind (\rits{yellow line}), and traces multiple kinematic components.

The above mentioned (see Section\ref{subsec:entrainment}) faster entrainment of cold cloud in galactic winds due to enhanced mass (and thus momentum) transfer from hot to cold medium is also an interesting avenue for observations. Specifically, we see a shorter entrainment time by a factor of 10, i.e., here $t_{\rm entrain}\ll t_{\rm drag}$ meaning \rits{clouds} are in the fully `momentum transfer' acceleration regime \citep[see, also][]{Gronke2020,Tonnesen2021ApJ...911...68T} and ram pressure is negligible. 

Fig. ~\ref{fig:Drag time} shows the `entrainment distance', \rits{defined as the distance at which the relative velocity between the cloud and wind reaches} $\Delta v / v_{\rm wind} \sim 0.1$, \rits{plotted against $1+\mathcal{M}_{\rm turb}$}. We find that this distance \rits{beyond} 
which 
the clouds are essentially comoving with the hot medium $d_{\rm ent}$ decreases as $\propto (1+\mathcal{M}_{\rm turb})^{-3}$.
In fact, we can convert these numerical findings to an expected observable velocity gradient along the tail \rits{$\delta v/\delta d$; where $\delta v$ is the velocity difference along a distance $\delta d$. As $d_{\rm ent}$ is the characteristic distance over which the cloud accelerates and essentially becomes comoving with the hot wind, the cloud velocity changes from 0 to nearly $v_{\rm wind}$, i.e., $\delta v \sim v_{\rm wind}$ and $\Delta d\sim d_{\rm ent}$. Therefore, the velocity gradient along the tail of the cloud
can be approximated as}
\begin{align}\label{eq:velocity_gradient}
    \frac{\delta v}{\delta d}&\sim t_{\rm drag}^{-1} (1+\mathcal{M}_{\rm turb})^{3}\\
    &\approx \rm 1\ km\,s^{-1}\,pc^{-1} \left(\frac{v_{\rm wind}}{300 \rm\ km\ s^{-1}}\right)
    \left(\frac{100}{\chi}\right)
    \left(\frac{{\rm 3\, pc}}{R_{\rm cl}}\right)
    \left(\frac{1+\mathcal{M}_{\rm turb}}{1.3}\right)^{3},\nonumber
\end{align}
for a surviving cloud in typical wind conditions.
Naturally, this acceleration gradient would be present only until the cloud reaches the wind velocity, i.e., at distances larger than $d_{\rm ent}$ the cold gas velocity approaches $\sim v_{\rm wind}$.
This implies that by observing the distance where the velocity-distance relation becomes flat, one can estimate the turbulence in the hot wind, for an assumed or measured cloud size and density contrast.
On the other hand, if one assumes that the turbulent energy is the same in the hot and cold phases (same $\mathcal{M}_{\rm turb}$ in both phases) or measured $\mathcal{M}_{\rm turb}$ by other means, one can estimate the size of the cold gas clouds $R_{\rm cl}$. 
The velocity gradient can be correlated to the observed luminosity (as discussed above) as well as the observed line widths, which are expected to be of order $\sim \mathcal{M}_{\rm turb}c_{\rm s,hot}$ (cf. Fig.~\ref{fig:cloud_MgII_columns}).

High-velocity clouds (HVCs), which are observed close to the Galactic disk, are also going to be affected by the turbulent nature of the CGM. These clouds provide a natural laboratory for studying the effects of turbulence on gas dynamics. While HVCs show a `droplet shape' \citep{
PutmanAnnurev2012, Westmeier2018,  Barger_2020}, they do generally not exhibit long tails. In simple `cloud crushing' simulations one finds tail lengths of order $\sim 0.2-0.5\,\chi\,R_{\rm cl}$ (\citealp{Gronke2020} but see \S~\ref{subsec:morphology_with_turb} in which we find pre-factors closer to unity)\footnote{Note that for infalling clouds one would expect even longer tails \citep{TanOhGronkeFallingClouds2023}.}, i.e., HVCs would easily be expected to span $\gtrsim 2\,$kpc. While photoionization of the gas might play a role in rendering (part of) the tail undetectable in $21$\,cm, such long tails seem inconsistent with current observations. 
Including turbulence in the hot medium might alleviate this tension. We find generally the tails to be a factor of few shorter (cf.~\S~\ref{subsec:turbulent_enhancement_of_cold_gas}), and more fragmented, thus, reducing self-shielding. 
A potential future interesting avenue is to use background quasars to probe the (ionized) environment around HVCs and to map out the exact tail morphology.

Another interesting laboratory for the here discussed multiphase gas dynamics are ram pressure stripped galaxies. As here the setup and involved processes are different to the `cloud crushing' simulations discussed here, we will explore the effect of extrinsic turbulence on ram pressure stripped galaxies in future work.











\subsection{Caveats and future work}
High-resolution simulations of astrophysical multiphase systems are essential for accurate comparisons with observations and reliable predictions, particularly those involving cold gas. 
Without sufficient spatial resolution to resolve the mixing layer in cloud-crushing simulations, the full multiphase gas structure cannot be thoroughly studied. However, while it is well-established that for numerical convergence of the phase structure, the turbulent mixing layer must be resolved by at least four cells (\citealt{Koyama+2004, Fielding2020ApJ}) or by including thermal conduction (\citealt{Bruggen2016}). 
Failure to do so can result in the gas cooling and fragmenting to the grid scale, leading to an overestimation of cold gas mass (or `overcooling'). 

However, in `cloud-crushing' studies that include radiative cooling, it is sufficient to resolve only the outer mixing scale $\sim r_{\rm cl}$ in order to obtain convergence in the cloud mass growth rate \citep{Gronke2020,Dutta2025,Abruzzo2022ApJ}. This is expected: for growing clouds $t_{\rm cool,mix}<t_{\rm cc}$, i.e., in this regime mixing rather than cooling sets the bottleneck. Since the mixing time increases with scale, the largest eddies regulate the cascade and thereby $\dot m$, which motivates the above resolution criterion \citep[see][]{TanOhGronke2021, Prateek2025arXiv}.\footnote{Note that a convergence in $\dot m$ does not imply the individual 
cooling layers are resolved and convergence, e.g., in the amount of mixed gas, is found.}

In determining the interplay between shear and external turbulent forcing on cloud survival, we have ignored several additional physical processes. 
One such process is thermal conduction, which can suppress hydrodynamic instabilities and potentially alter the morphology of the surviving clouds.
However, for typical conditions of the CGM, large enough clouds are difficult to evaporate(\citealt{Bruggen2016, Armillotta2017, LiHopkins2020}). Moreover, turbulent diffusion can predominantly mediate heat transport over thermal conduction (refer to \citealt{TanOhGronke2021} for a detailed explanation). For these reasons, excluding thermal conduction would not significantly impact our conclusions, particularly concerning the cold mass growth rate $\dot{M}_{\rm cl}$.

We neglect the presence of magnetic fields, which can influence cloud survival due to magnetic draping \citep{Dursi2008,McCourt2015} -- in particularly in combination with radiative cooling \citep{Fernando2023} -- as well as their overall morphology \citep{Shin2008ApJ,Gronnow2018}. 
Nevertheless, the main conclusions of our study, particularly concerning the mass growth of clouds and turbulent mixing, are expected to remain largely unaffected. While magnetic fields can suppress hydrodynamic Kelvin-Helmholtz and Rayleigh-Taylor instabilities at small scales \citep{Chandrasekhar1981}, the dominant driver of cloud growth arises from instabilities at the cloud scale which are resolved in our study \citep{Gronke2020,Fernando2023}. In particular, \citet{Das2024MNRAS} showed that in the case of larger scale extrinsic turbulence, the mass transfer rate is unaffected by magnetic fields.

We also neglect viscosity. Including viscosity can be particularly  important in the evolution of the large-scale turbulence apart from its effect on reducing hydrodynamic mixing and redistribution of energy among the phases. 
One might expect that viscosity also affects the mass transfer rate between the phases ($\dot M$) since it affects the turbulent cascade and, thus, the mixing process (see, e.g., discussion in \citealp{TanOhGronke2021}). However, recently \citet{Tirso2025arXiv} showed in turbulent mixing layer simulations that in the strong cooling regime, cooling sharpens the density contrast, thus, compensating the effect of viscosity and leaving $\dot M$ nearly unaffected. 

The spatial geometry and structure of the wind are expected to strongly influence the evolution of cold gas. Recently, \citet{Dutta2025} showed that the growth of cold gas is suppressed when clouds traverse an adiabatically expanding wind. Far from the source of the outflow, the turbulence might weaken and saturate. Since we demonstrate that the cloud evolution depends sensitively on the turbulent Mach number, an interplay between mass growth in presence of external turbulence and a suppression from adiabatic expansion of the wind, could result in diverse evolution scenarios, which we plan to explore.  
In realistic outflows, further complexity arises from the interaction of multiple clouds uplifted together \citep{Seidl2025arXiv}, as well as their collective back-reaction on the wind \citep{FieldingBryan2022}. Connecting different wind profiles and cloud populations within a turbulent background is therefore an interesting direction for future work.

Another important aspect neglected in our study is the impact of heating, and in particular how a global thermal (in-)balance affects the mass transfer between the phases. In this work, the simulation domain was merely heated through turbulence. However, in reality, other sources of heating -- such as radiation or cosmic rays exist. Global thermodynamics can have an important influence on long term fate of multiphase gas \citep{Sharma2010, Mohapatra2022MNRAS}.
A thorough exploration of how these modifications to 
the (effective) cooling curve alter the outcome of the simulation is beyond the scope of this study, but we plan to revisit this point in future work.







\section{Conclusions}\label{sec:conclusions}
The survival and entrainment of cold clouds in their hostile environment 
is a long-standing issue in galaxy evolution. Multiphase outflows are a combination of laminar and turbulent flows. Despite their interconnected nature, simulations often model the interaction of cloud with a laminar wind, and that with a turbulent flow separately. In this work, we have explored the survival of clouds in galactic outflows in the presence of externally driven turbulence, \rits{by introducing continuous turbulent forcing into the classic cloud-crushing setup}. 
The key findings of our work are:

\begin{enumerate}
    \item \textit{Interplay between cooling and turbulence}: 
    \rits{In the classic cloud-crushing scenario, a} dense cloud facing a laminar wind is constantly mixed, producing a lot of intermediate-temperature gas. \rits{The cloud is typically destroyed on the cloud-crushing timescale $t_{\rm cc}$, unless cooling is rapid enough to condense the mixed gas back into a cold phase ($t_{\rm cool, mix}<t_{\rm cc}$).
    Our findings reveal that despite the presence of additional turbulence, cold clouds can survive if the mixed-phase gas cools sufficiently quickly. Specifically, the survival criterion is only slightly modified (see Fig. \ref{fig:overview_cc_turb}): cold clouds can survive if the cooling time of the mixed gas $t_{\rm cool, mix}$ is shorter than a modified destruction timescale $\tilde{t}_{\rm cc}$, i.e., $t_{\rm cool,mix}/\tilde{t}_{\rm cc}<1$ where, 
    \[\tilde{t}_{\rm cc} = 
    \frac{t_{\rm cc}}{\sqrt{(1+\left(\mathcal{M}_{\rm turb}/\left(f_{\rm mix}\mathcal{M}_{\rm wind}\right)\right)^2)}}.\]
    Here, $\mathcal{M}_{\rm turb}$ is the rms turbulent Mach number of the hot wind, $\mathcal{M}_{\rm wind}$ is its Mach number for the bulk flow, and $f_{\rm mix}\sim 0.6$ is a fudge factor.
    Thus, even in turbulent environments, efficient cooling enables clouds to survive and grow.}

    \item \textit{Enhanced cloud growth with turbulent mixing}: At short cooling timescales  ($t_{\rm cool, mix}/t_{\rm cc}\lesssim 10^{-2}$), increasing the turbulent Mach number $\mathcal{M}_{\rm turb}$ in the hot wind leads to a significant increase in the cold gas mass \rits{-- upto an order of magnitude (see Fig. \ref{fig:mass_growth}).} 
    This \rits{cold gas mass} growth can be attributed to turbulence \rits{stretching the cold gas such that the area of the interface where hot and cold gas can mix and cool efficiently is enhanced. This is  
    seen in Fig. \ref{fig:mass_area_growth}, where the increase in the cold gas mass growth rate $\dot{M}_{\rm cl}$ , is corresponded by a comparable increase in the area (of an iso-surface at $\sim 2T_{\rm cl}$).}
    Interestingly, the derived mixing velocity $v_{\rm mix}$ remains relatively unaffected \rits{(see Fig. \ref{fig:mixed_velocity})}.
    
    \item \textit{Faster entrainment of cold gas with turbulence}: As a consequence of fast growth of cold gas with increasing turbulence, we find that clouds are entrained in a turbulent wind by efficient `momentum transfer' (see Fig. \ref{fig:correlation_massgrowth_entrainment} where a strong correlation between mass enhancement time and entrainment time). This leads also to a larger velocity gradient in galactic winds which is potentially observable (cf. Eq.~\eqref{eq:velocity_gradient}).

    \item \textit{Suppressed cloud growth with strong turbulence}: Turbulence is also seen to suppress the cold mass growth in cases with relatively weak cooling ($1\gtrsim t_{\rm cool,mix}/t_{\rm cc} \gtrsim 0.3$) and strong turbulence ($\mathcal{M}_{\rm turbl}\gtrsim 0.5)$. In extreme cases, this can lead to the eventual (accelerated with respect to the laminar car) destruction of cold gas. 

    \item \textit{Growth rate and turbulent Mach number}: In scenarios where cold mass growth is evident, we observe a strong dependence of the growth rate of cold mass $\dot{M}_{\rm cl}$ on the turbulent Mach number $\mathcal{M}_{\rm turb}$. Fig. \ref{fig:overview_growth_rate} suggests that $\dot{M}_{\rm cl} \propto \mathcal{M}_{\rm turb}^{3/2}$. This is also reflected in the area available for intermediate temperature which grows with the geometric criterion and is proportional to $M_{\rm cl}^2/3$. 

    \item \textit{Cloud morphology}: Unlike the elongated cold tails formed in laminar winds, we find that turbulent forcing results in morphologically distinct- \rits{clumpy and irregular} 
    cold gas structures, particularly during the later stages of cloud crushing. 
    \rits{Consequently, long filamentary tails are unlikely to persist to large distances in galactic outflows.} 
    We calculate the MgII 2796 $\mathring{A}$ equivalent width along random sightlines through our simulation domain with \texttt{AstroPlasma}, which uses pre-computed \texttt{Cloudy} models, assuming \rits{photo+collisional ionization equilibrium. }
    The presence of turbulence results in broader absorption profiles that trace multiple kinematic components in the tails (cf. Fig. \ref{fig:cloud_MgII_columns}). 
\end{enumerate}


Overall, our findings highlight the critical role of turbulence in the dynamics of multiphase gas in galactic environments and its strong influence on the survival and morphological evolution of cold gas.

\section*{Acknowledgements}
The simulations were carried out at the HPC cluster Freya at Max Planck Computing and Data Facility (MPCDF). RG acknowledges Rajsekhar Mohapatra for useful discussions. 
MG thanks the Max Planck Society for support through the Max Planck Research Group, and the European Union for support through ERC-2024-STG 101165038 (ReMMU).
This research was supported by the International Space Science Institute (ISSI) in Bern, through ISSI International Team project \#545. This work was supported by the Programme National PNST of CNRS/INSU co-funded by CNES and CEA.

\section*{Data Availability}
Data related to this work will be shared on reasonable request to the
corresponding author. All the simulation, analysis, and visualization codes used in this work are hosted on the following \texttt{GitHub} repository -- \url{https://github.com/RitaliG/cc_driven_turbulence.git}. 



\bibliographystyle{mnras}
\bibliography{turbRadCC} 




\appendix


\section{The derived mixing velocity}
\label{app:vmix_derived}
\begin{figure}
    \centering
    \includegraphics[width=\columnwidth]{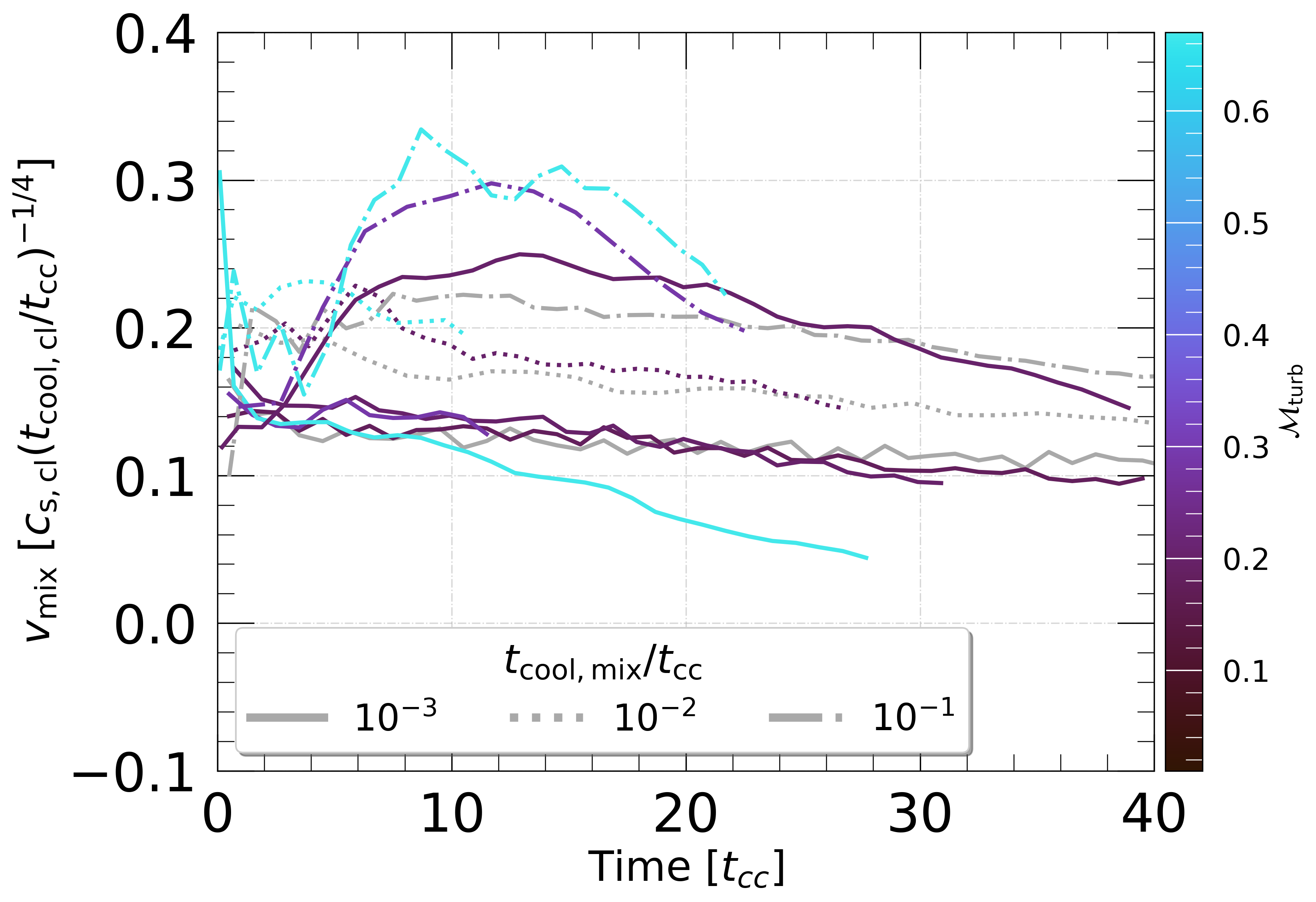}
    \caption{The derived mixing velocity $v_{\rm mix}$ as a function of time in units of cloud crushing time $t_{\rm cc}$. We normalize $v_{\rm mix}$ with ($c_{\rm s,cl}(t_{\rm cool, cl}/t_{\rm sc})^{-1/4}$), \rits{where $c_{\rm s,cl}$ is the sound crossing time in the cloud, $t_{\rm cool, cl}/t_{\rm cc}$ is the ratio of the cooling time of the cloud and the `standard' cloud crushing time. The presence of turbulence does not significantly alter $v_{\rm mix}$; over time, the slight decrease in $v_{\rm mix}$ is mainly due to the rapid growth of the cloud surface area $A_{\rm turb}$.
    }
    }
    \label{fig:vmix_evolution}
\end{figure}

The derived mixing velocity $v_{\rm mix}$ as a function of time in units of cloud crushing time $t_{\rm cc}$ is shown in Fig. \ref{fig:vmix_evolution}. We normalize $v_{\rm mix}$ with ($c_{\rm s,cl}(t_{\rm cool, cl}/t_{\rm cc})^{-1/4}$), where $c_{\rm s,cl}$ is the sound crossing time in the cloud, $t_{\rm cool, cl}/t_{\rm cc}$ is the ratio of the cooling time of the cloud and the `standard' cloud crushing time.
In the standard cloud crushing simulations, $v_{\rm mix}$ typically reaches $\sim 0.2$ in these units and, we find that turbulence does not significantly alter this value. Over time, $v_{\rm mix}$ has a slightly decreasing value, which is subject to the fact that the area grows rapidly and the derived velocity -- being inversely related to area -- can be affected numerically.

\section{Cloud destruction with externally driven turbulence}
\begin{figure}
    \centering
    \includegraphics[width=\columnwidth]{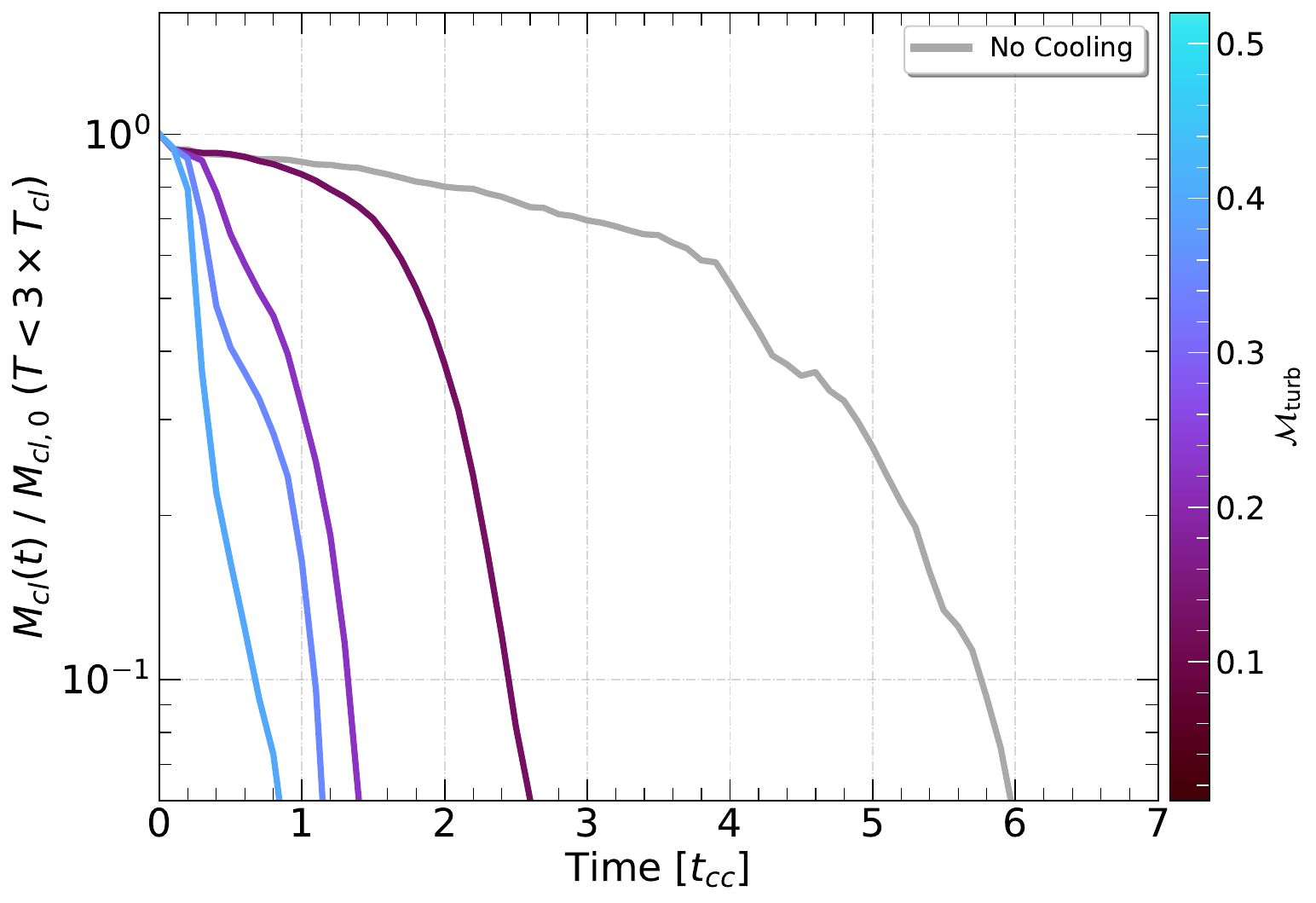}
    \caption{\rits{Evolution of cold gas mass in adiabatic cloud crushing simulations without radiative cooling. The gray line is without turbulence while lines in color are according to the increasing strength of turbulent forcing (as indicated in the colormap). Stronger turbulence enhances mixing at the cloud–wind interface, leading to faster cloud destruction compared to the laminar case.}
    }
    \label{fig:mass_evol_adiabatic}
\end{figure}

Fig. \ref{fig:mass_evol_adiabatic} shows the mass of cold gas in adiabatic cloud crushing simulations with increasing turbulent forcing, as indicated in the colorbar. The gray line corresponds to the run without turbulence, magenta to the lowest turbulent Mach number in the hot phase $\mathcal{M}_{\rm turb}\sim 0.2$, and cyan to the highest $\mathcal{M}_{\rm turb}\sim0.7$. As turbulence grows, the shear at the cloud-wind interface becomes stronger, enhancing the mixing of the cold cloud into the hot wind. In the absence of radiative cooling, this  efficient mixing can be accompanied by additional heating of the cloud due to turbulence, accelerating its evaporation. As a result, without cooling, turbulence significantly shortens the survival time of cold clouds in a hot turbulent wind.

    

\section{Differential emission}\label{app:differential emission}
\begin{figure*}
    \includegraphics[width=0.3\textwidth]{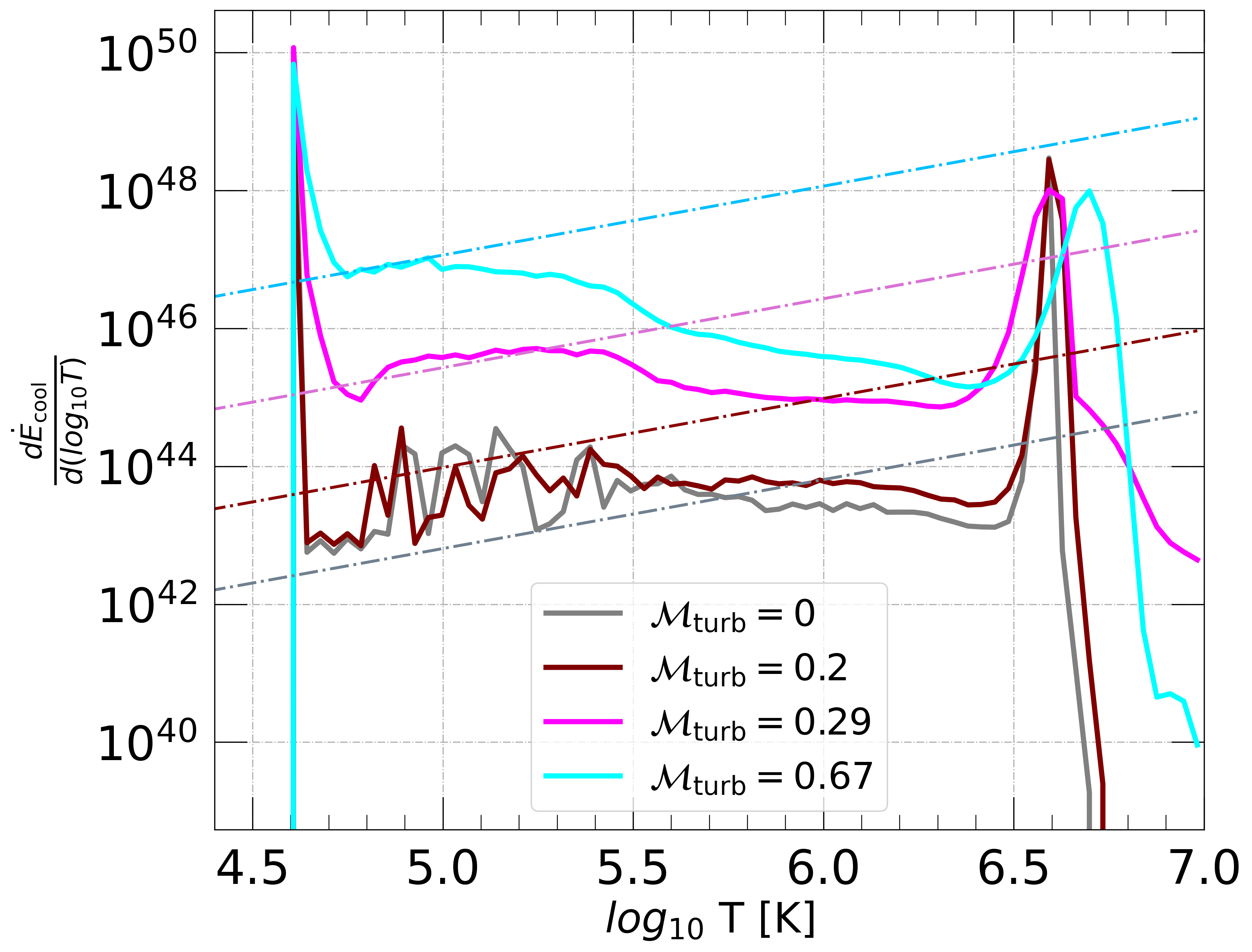} 
    \includegraphics[width=0.3\textwidth]{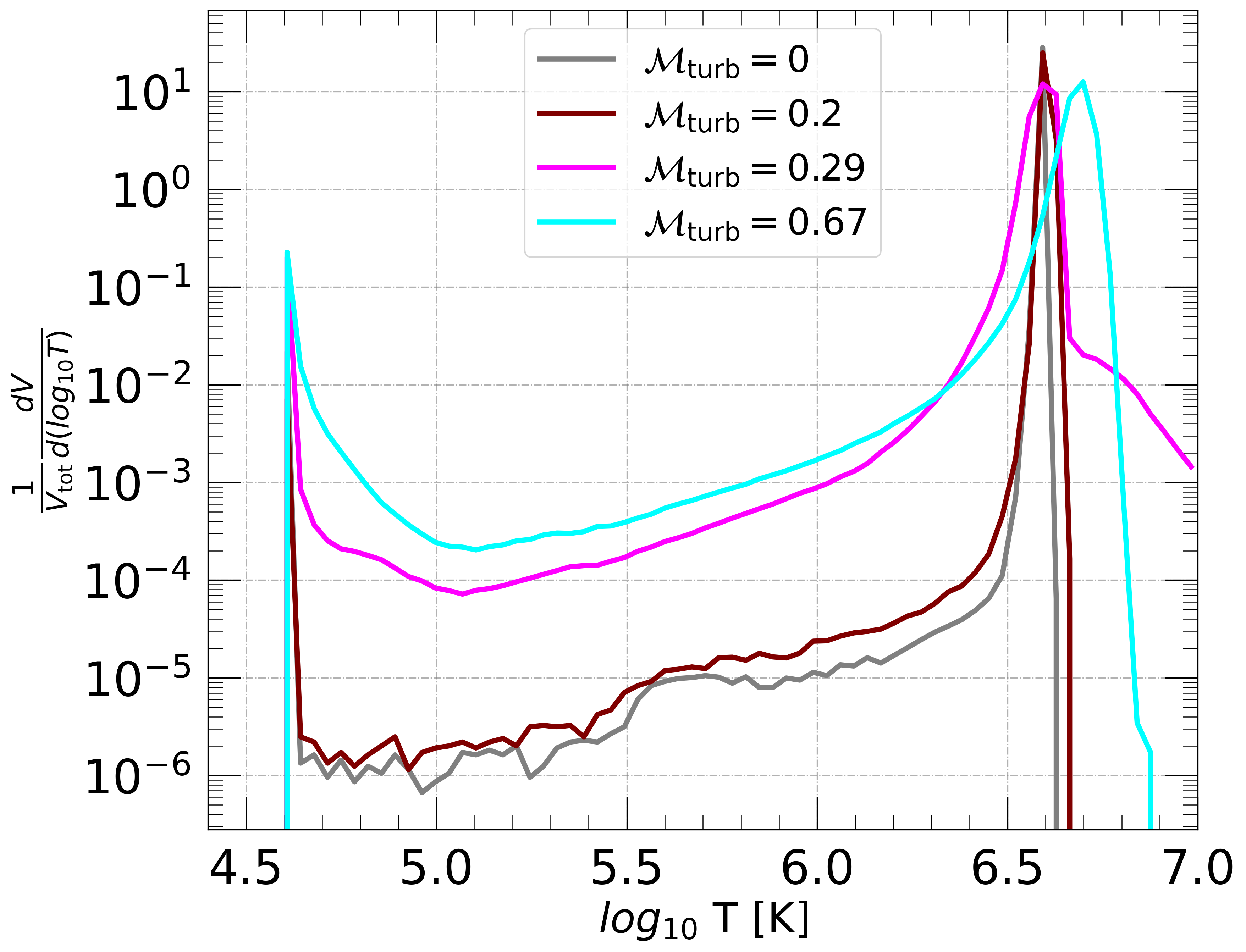}
    \includegraphics[width=0.3\textwidth]{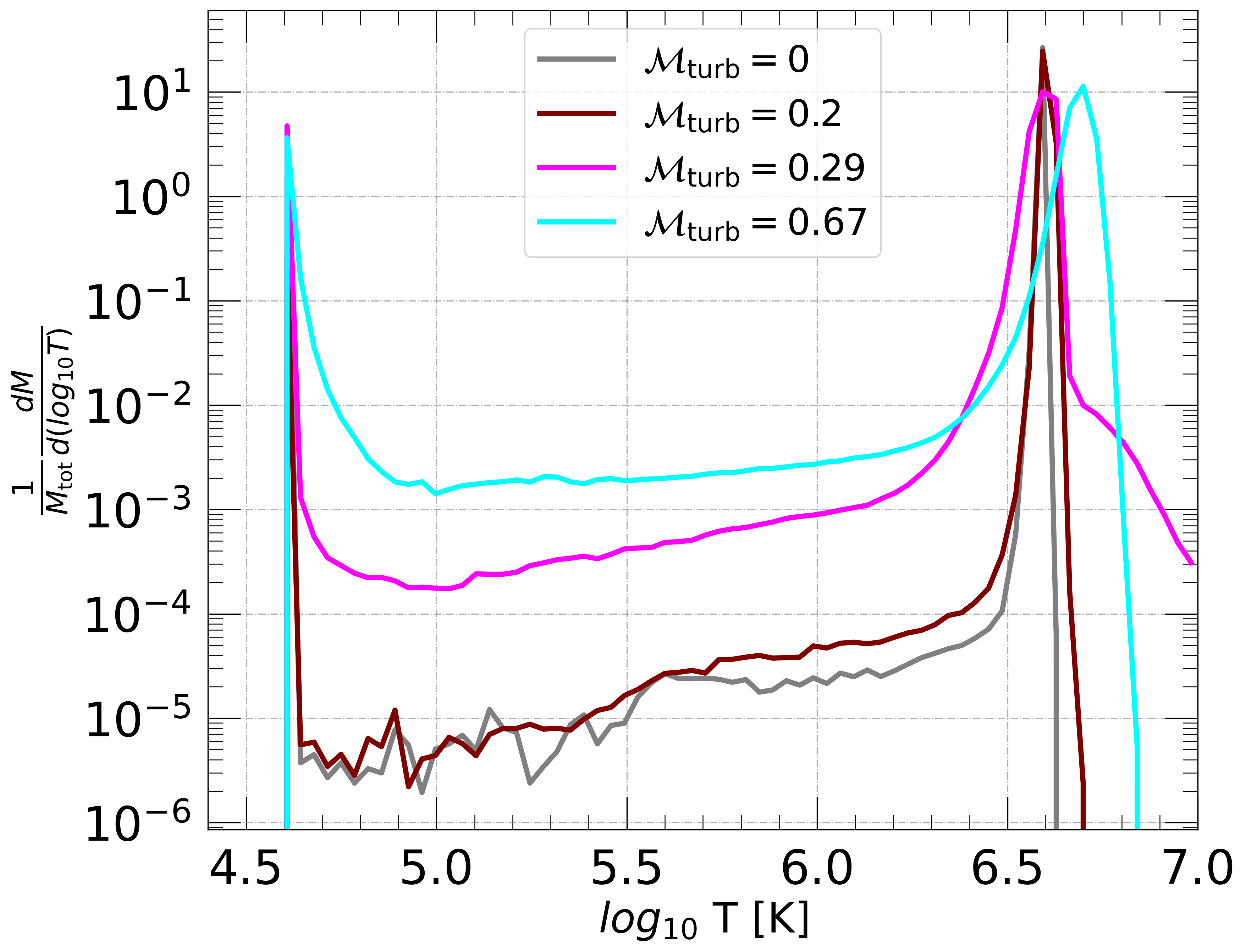}
    \caption{\emph{Left panel:} The differential emission as a function of temperature, for various turbulent Mach number, at a time aroiund $5t_{\rm cc}$ after the cloud is entrained ($t_{\rm ent}$ in Fig. \ref{fig:correlation_massgrowth_entrainment}).
    \emph{Middle and right panels:}The volume-weighted and mass-weighted PDF temperature at various Mach numbers (at the same time as the top panel) 
    }
    \label{fig:differential_emission}
\end{figure*}

\rits{
The middle and right panels of Fig. \ref{fig:differential_emission} show the distributions in volume and mass (respectively) of gas at different phases from our simulations. Each curve corresponds to a simulation with a different turbulent Mach number, as indicated by the legend accompanying the figure. The Mach number of the wind is set to $1.5$ -- unchanged across all the simulations. Simulation snapshots at a time of around $5t_{\rm cc}$ is chosen for the analysis. In line with the enhancement of cold mass growth due to turbulence (cf. Fig. \ref{fig:mass_growth}), the phase distributions in a turbulent wind show a significantly increased volume and mass of the intermediate temperature (warm) gas, which can cool very efficiently leading to a higher peak in the cold gas mass and volume. For the hot phase, turbulence results in broadening of the hot gas peak as well as a small shift to an even hotter temperature, which can be attributed to turbulent heating. Overall, the enhanced mass growth of clouds with turbulence is due to enhanced turbulent mixing as well as strong local cooling flows around the clouds, which we elaborate further.

The left panel of Fig. \ref{fig:differential_emission} shows the differential emission at different gas phases from our simulations. For calculating the differential emission, we have chosen a simulation snapshot (for each turbulent Mach number) at a time of around $5t_{\rm cc}$ when the cloud is well entrained (cf. $t_{\rm ent}$ in Fig. \ref{fig:correlation_massgrowth_entrainment}). Once entrained, we find tentative evidence that the cloud develops a steady subsonic inflow of mass from the intermediate temperature phase to the cold phase in accordance with the "local cooling flow" solutions (\citealt{Dutta2022}). Such local cooling flows that continuously feed cold gas into the cloud (co-moving with the wind) are a natural outcome of advection in response to the pressure gradient between the hot and the cold gas that develops across the boundary layers due to radiative cooling in these layers.\footnote{A small pressure is sufficient to kickstart the flow, which arises due to fast radiative cooling of the intermediate temperature gas in the boundary layer.} \citealt{Dutta2022} (see also \citealt{Kanjilal2021}) find that the differential emission $d\dot{E}_{\rm cool}/d(\text{log}_{10}T)$, in case of a pure steady cooling flow, follows a simple power law scaling with temperature of the gas and mass inflow rate, namely
\begin{equation}
    \label{eq:cooling_diff_emm}
    \frac{d\dot{E}_{\rm cool}}{d(\text{log}_{10}T)} \propto  \left(\frac{T}{10^{5.5 } K} \right)\left(\frac{\dot{M}}{10^{-4} M_{\odot}\text{yr}^{-1}} \right).
\end{equation}

We find that this relation holds for the efficiently cooling warm phase which predominantly populates the boundary layer between the cloud and the wind. The differential emission stays close to the cooling flow relation even with moderately strong turbulence (cf. magenta line; $\mathcal{M}_{\rm turb}=0.29$ in Fig. \ref{fig:differential_emission}). However, at a higher turbulent forcing ($\mathcal{M}_{\rm turb}\sim 0.7$), we find deviation from the simple cooling flow solution. This is because, in this regime, strong turbulence is expected to dominate the energy transport across gas at different phases due to turbulent conduction, which can no longer be neglected (\citealt{TanOh2021}). Additional energy transport by turbulent conduction limits the phases where we find a cooling flow (Eq. \ref{eq:cooling_diff_emm}) to a narrow range in temperature at $\sim 10^5$ K. Therefore, in addition to advection and radiative cooling, turbulent conduction is also an important physical effect in `real' outflows. Investigating the nature of local cooling flow solutions and their associated differential emission in presence of turbulent conduction can reveal different interesting and analytically tractable physical regimes of cold mass growth in clouds -- beyond the scope of the present study. 
}





\bsp	
\label{lastpage}
\end{document}